\documentclass[12pt]{article}
\usepackage[margin=0.95 in]{geometry}
\usepackage{amsmath} 
\usepackage{amssymb,amsfonts}
\usepackage[all]{xy}
\usepackage{graphicx}
\usepackage{simplewick}
\usepackage{amsmath}
\usepackage{amssymb}
\usepackage{float}
\usepackage{array}
\usepackage{tikz}
\usepackage{mathtools}
\usepackage{mathrsfs} 
\usepackage[hidelinks]{hyperref}
\usepackage{cite}
\numberwithin{equation}{section}
\usepackage[utf8x]{inputenc}

\newcommand{\be}{\begin{equation}}
\newcommand{\ee}{\end{equation}}
\def\bea{\begin{eqnarray}}
\def\eea{\end{eqnarray}}

\newcommand{\taust}{{\tau}_{\text{s.t.} }}

\newcommand{\ii}{i}

\DeclareMathOperator{\Tr}{Tr}

\numberwithin{equation}{section}
\numberwithin{table}{section}

\begin{document}

\hypersetup{pageanchor=false}
\begin{titlepage}
\vbox{
    \halign{#\hfil         \cr
           } 
      }  
\vspace*{15mm}
\begin{center}
{\Large \bf Long Strings 
and 
Quasinormal Winding Modes}

\vspace*{10mm} 

{\large Sujay K. Ashok$^{a,b}$ and Jan Troost$^c$}
\vspace*{8mm}

$^a$The Institute of Mathematical Sciences, \\
		 IV Cross Road, C.I.T. Campus, \\
	 Taramani, Chennai, India 600113

\vspace{.6cm}

$^b$Homi Bhabha National Institute,\\ 
Training School Complex, Anushakti Nagar, \\
Mumbai, India 400094

\vspace{.6cm}

$^c$Laboratoire de Physique de l'\'Ecole Normale Sup\'erieure \\ 
 \hskip -.05cm
 CNRS, ENS, Universit\'e PSL,  Sorbonne Universit\'e, \\
 Universit\'e  Paris Cit\'e 
 \hskip -.05cm F-75005 Paris, France	 

\vspace*{0.8cm}
\end{center}

\begin{abstract}
We compute the path integral for a particle on the covering group of SL$(2,\mathbb{R})$ using a decomposition of the Lie algebra into adjoint orbits. We thus intuitively derive the Hilbert space of the particle on the group including discrete and continuous representations. Next,  we perform a Lorentzian hyperbolic orbifold of the partition function and relate it to the Euclidean BTZ partition function.  
We use the particle model to inform further discussion of the spectral content of the one loop vacuum amplitude for strings on BTZ black hole backgrounds. We argue that the poles in the loop integrand  code  contributions of long string modes that wind the black hole. We moreover identify saddle point contributions of quasinormal winding modes. 

\end{abstract}

\end{titlepage}
\hypersetup{pageanchor=true}

\setcounter{tocdepth}{2}
\tableofcontents

\section{Introduction}
String theory is a unitary theory of quantum gravity. The theory has black hole solutions. It would be of great interest to show that non-perturbative fundamental string theory exhibits a unitary evolution in the presence of a black hole. At present, we are far from attaining this unification grail of theoretical physics.

A more modest goal is to study perturbative scattering amplitudes on a black hole background. A prime candidate geometry to ponder is the asymptotically $AdS_3$ BTZ black hole \cite{Banados:1992wn,Banados:1992gq}. It has an asymptotic structure that allows for holography on the one hand, and for a deceptively simple algebraic world sheet description on the other. Indeed, the  BTZ black hole background with Neveu-Schwarz Neveu-Schwarz flux is described by an orbifold of a Wess-Zumino-Witten model on the universal cover $\widetilde{G}$ of the group $G=SL(2,\mathbb{R})$ \cite{Banados:1992gq,Natsuume:1996ij}. As such, we have multiple tools at our disposal for studying strings on this black hole background. We may hope to make progress then in the modest study of perturbative string propagation on a fixed black hole geometry. To start with, we can attempt  a  description of the complete spectrum of fundamental strings propagating on the Lorentzian BTZ geometry. Steps towards this goal were taken in \cite{Natsuume:1996ij,Hemming:2001we,Troost:2002wk,Hemming:2002kd,Rangamani:2007fz,Berkooz:2007fe,Mertens:2015ola,Ashok:2021ffx,Nippanikar:2021skr}. The determination of the full spectrum is a recognized and hard open problem. 

In this paper, we approach this study in two distinct manners. Firstly, we consider the particle limit of the problem directly in the Lorentzian geometry. We study the partition function of a particle in a BTZ black hole background, arising as a hyperbolic orbifold of the covering group $\widetilde{G}$.  We  recall how to derive the Hilbert space of a particle on a compact group from a path integral \cite{Alekseev:1988vx,Alekseev:1990mp}. We then show how to extend this reasoning to particles on the non-compact group SL$(2,\mathbb{R})$ as well as its universal cover $\widetilde{G}$, exploiting the coadjoint orbit quantization results from \cite{Ashok:2022thd}. This provides us with a direct derivation and interpretation of the Hilbert space for a particle on $AdS_3$ as well as the partition function twisted by global symmetry group elements. In a side step, we show that the elliptic orbifold of $AdS_3$ in the Lorentzian neatly connects with a particle partition function on Euclidean $AdS_3$. 
Next, we analyze the Lorentzian hyperbolic orbifold of the BTZ partition function and stress the differences with the elliptic case.
The main difference lies in having either a topologically trivial or non-trivial thermal circle direction. 

In a second part of the paper, we revisit the  partition function for strings on the Euclidean BTZ black hole background. While the partition function is partly understood \cite{Maldacena:2000kv,Berkooz:2007fe,Mertens:2015ola,Ashok:2021ffx,Nippanikar:2021skr}, we push our understanding further still. In particular, extending the intuition developed in finite temperature $AdS_3$ \cite{Maldacena:2000kv}, we identify pole regions in the Euclidean BTZ partition sum that correspond to the contribution of long string winding modes that surround the black hole. We also  identify short winding string contributions in a saddle point approximation, and associate them to quasinormal winding modes. We moreover are able to identify a bound on their sl$(2,\mathbb{R})$ spin from this analysis. 

The paper is partitioned as follows. 
It starts with the analysis of path integrals describing a particle on a group manifold in section \ref{ParticleOnGroup}. Using our knowledge of twisted partition functions on the group, we formulate the partition functions for particles on orbifolds of groups in section \ref{ParticleOnOrbifold} and discuss how a Lorentzian orbifold partition sum can be analytically continued to a Euclidean partition function. Sections \ref{ParticleOnGroup} and \ref{ParticleOnOrbifold} are a sequel to
\cite{Ashok:2022thd}. In section \ref{Strings}, we identify the contribution of Lorentzian long strings to the Euclidean partition function of string theory on the BTZ hyperbolic orbifold of $AdS_3$. We lay bare the multi-particle nature of the contribution explicitly and provide an argument and description of which poles in the partition function can be identified with physical long string contributions. We also identify quasinormal winding mode contributions. This second part is a sequel to \cite{Ashok:2021ffx}.  Both parts make progress in the task  to rhyme the Euclidean partition function with the Lorentzian spectrum of perturbative string theory in the NSNS BTZ black hole background.  We conclude in section \ref{Conclusions} by summarizing our main results and pointing out a few related open problems.

\section{The Path Integral for a Particle on a Group}
\label{ParticleOnGroup}
In this section, we recall how to derive the Hilbert space for a particle on a compact group directly from path integration \cite{Alekseev:1988vx,Alekseev:1990mp}. The technique is to show the equivalence of the path integral to an integration over left and right group elements that parameterize coadjoint orbits, then to apply coadjoint orbit theory \cite{Kirillov}. We  extend this transparent viewpoint on the derivation of the particle Hilbert space to a non-compact group. The main new feature is the split of the set of Lie algebra elements into intrinsically distinct classes.

\subsection{The Particle on a Group}
A particle on a group $G$ has a phase space which is the cotangent bundle $T^\ast G$ to the group. 
The wave-functions are  taken to be  quadratically integrable functions on the group $G$. The Hilbert space is ${\cal L}^2(G)$. The latter allows for a description in terms of left and right regular actions of the group on itself and as such decomposes in a direct sum of left and right irreducible representations. Firstly, we wish to make this canonical mathematical statement transparent from a path integral point of view.

To that end, we briefly review the approach of  \cite{Alekseev:1990mp}, which in turn is based on the path integral description of coadjoint orbits provided in \cite{Alekseev:1988vx}. The cotangent bundle $T^\ast G$ of a group is a symplectic manifold. We define fields $t: L \rightarrow \mathfrak{g}^\ast$ which are maps from the world line $L$ of the particle into  the dual of the Lie algebra $\mathfrak{g}^\ast$ as well as the group valued field $g: L \rightarrow G$. The canonical symplectic form $\Omega$ on the phase space $T^\ast G$ is \cite{Alekseev:1990mp}:
\begin{equation}
\Omega = \frac{1}{2} \text{Tr} (dt dg g^{-1} + t dg g^{-1} dg g^{-1}) =  d \left(\frac{1}{2} \text{Tr} t dg g^{-1} \right) \, .
\end{equation}
We can view the dual Lie algebra variable $t$ as corresponding to momenta and the group variable $g$ as coordinates.  
We quantize trajectories in the cotangent bundle with 
an action which is a primitive of the symplectic form $\Omega$:
\begin{equation}
S = \frac{1}{2} \int_{L} \ast \text{Tr} ( t dg g^{-1}) \, ,
\end{equation}
where $\ast$ indicates the pullback to the world line $L$. 
The path integral is:\footnote{We have the possibility of adding a quadratic Casimir Hamiltonian $H=\frac{1}{R^2} \text{Tr} t^2$. If we integrate out the variable $t$ then, we obtain a particle model on the group with a second order kinetic term. This model has a Hamiltonian equal to the Laplacian on the group $G$.}
\begin{equation}
Z = \int [d t(\tau)] [dg(\tau)] e^{i S[t,g]}  \, .
\end{equation}
The model has both a right and left global group invariance. It permits a left and right group $G_l \times G_r$  action  and there are conserved charges corresponding to these group actions. 
Below, our first aim is to compute a partition function of the type:
\begin{equation}
Z (\gamma_j^{l,r}) = \int [d t(\tau)] [dg(\tau)]_{\text{periodic}} e^{i S[t,g] + i \sum_j (\gamma_j^r H^j_r +\gamma_j^l H^j_l)} 
\label{HamiltonianInsertions}
%= \sum_\lambda \chi^l_\lambda (\gamma_j^l) \chi^r_\lambda(\gamma_j^r) 
\, ,
\end{equation}
where we twist the partition function with the insertion of commuting left and right  Hamiltonians $H^j_{l,r}$ that correspond to Cartan generators in the Lie algebras $\mathfrak{g}_l \oplus \mathfrak{g}_r$ of the symmetry group $G_l \times G_r$. The integration is over periodic maps such that the result has an interpretation as a trace.
Following \cite{Alekseev:1990mp}, we
can reparameterize the dual Lie algebra element as $t=g_l t_d g_l^{-1}$ in terms of a Cartan element $t_d$ and an element in the group $g_l$.
We find the action:
\begin{align}
S 
%&= \frac{1}{2} \int \text{Tr} (t_d d (f^{-1} g) (f^{-1} g)^{-1} + t_d f^{-1} df)
%\nonumber \\
 &= \frac{1}{2} \int \text{Tr} (t_d d g_r g_r^{-1} + t_d g_l^{-1} dg_l) \, ,
\end{align}
where $g_r=g_l^{-1} g$.
If we multiply $g_l$ by a Cartan element, $t_d$ is invariant, and we therefore  have a redundant parameterization and a gauge symmetry \cite{Alekseev:1990mp}. The global symmetries act on $g_r$ on the right and on $g_l$ on the left.  The local symmetry group conjugating $t_d$ contains the Weyl subgroup. 
The idea in \cite{Alekseev:1990mp} is to read the resulting path integral as a product of path integrals over co-adjoint orbits, both of a type set by the Cartan Lie algebra element $t_d$. Co-adjoint orbit quantization \cite{Kirillov,Alekseev:1988vx} then gives rise to the Hilbert space:
\be {\cal H} = \sum_i {\cal H}_i^l \otimes {\cal H}_i^r \, ,
\ee where we have a left-right correlation of the sum over representations $i$ due to the fact that the Cartan element $t_d$ appears in the action for the group valued field $g_r$ as well as the field $g_l$.  The path integrals are computed explicitly in Darboux variables  for the classical compact simple groups in \cite{Alekseev:1988vx} using a classical configuration space counterpart to the Gelfand-Tsetlin basis for classical Lie algebras. Since there are  differences in the treatment of the path integral compared to the ordinary quantization of co-adjoint orbits, and since the  details in our treatment differ from those of \cite{Alekseev:1988vx,Alekseev:1990mp}, we add a few pointers as to how this works for the compact group $G=SU(2)$ before turning to a generalization for non-compact groups. We refer the reader to e.g.  \cite{Alekseev:1988vx,Alekseev:1990mp,Kirillov} for the necessary background and the detailed derivation of the final formula.

\subsection{\texorpdfstring{The Example of  the Group $SU(2)$}{The Example of  the Group SU(2)} }
We concretely parameterize the path integral for the case of the simple compact group $G=SU(2)$.
We parameterize the group element  $g_l(\tau)$ as:
\begin{equation}
g_l = \left( \begin{array}{cc}
\cos \frac{\theta}{2} e^{i \frac{\phi+\psi}{2}} & i \sin \frac{\theta}{2} e^{i \frac{\phi-\psi}{2}} \\
i \sin \frac{\theta}{2} e^{i \frac{\psi-\phi}{2}} & \cos \frac{\theta}{2} e^{-i \frac{\phi+\psi}{2}} 
\end{array} \right) \, .
\end{equation}
The dual Lie algebra element $t$ we diagonalize such that it becomes proportional to the $\sigma_3$ Pauli matrix:
\be
t = g_l t_3 \sigma_3 g_l^{-1} \, .
\ee
In terms of these variables the  measure on the (dual) Lie algebra becomes:
\be
d t =
t_3 dt_3 \wedge t_3 \sin \theta   \, d\phi \wedge d \theta 
\, .
\ee
Thus, our path integral reads concretely:
\begin{align}
Z(\cdots) %&= \int dt \int dg e^{iS} 
 %\\
&= \int dt_3 \int (t_3 \sin \theta   \,  d\phi \wedge d \theta )
\int  (t_3  d g_r)
e^{\frac{i}{2} \int d \tau Tr (t_3 \sigma_3 g_l^{-1} \partial_\tau g_l ) 
+\frac{i}{2} \int d \tau Tr (t_3 \sigma_3 \partial_\tau g_r g_r^{-1}  )} (\cdots) \nonumber \, 
\end{align}
where the dots indicate possible operator insertions. We have split the measure into factors that are naturally associated to the left group element $g_l$ and the right group element $g_r$.  The integration over $g_l$ and $g_r$ are standard in co-adjoint orbit theory \cite{Alekseev:1988vx,Alekseev:1990mp,Kirillov}. There are two important extra elements in the path integration though; firstly, the measure $d g_r$ contains an integration over the variable $\psi$, as it is defined to be
\begin{equation}
g_r = \left( \begin{array}{cc}
\cos \frac{\theta_r}{2} e^{i \frac{\psi+\phi_r}{2}} & i \sin \frac{\theta_r}{2} e^{i \frac{\psi-\phi_r}{2}} \\
i \sin \frac{\theta_r}{2} e^{i \frac{\phi_r-\psi}{2}} & \cos \frac{\theta_r}{2} e^{-i \frac{\psi+\phi_r}{2}} 
\end{array} \right) \, .
\end{equation}
Note that $\psi$ also appears in $g_l$ (this happens via the field redefinition $g_r=g_l^{-1} g$). Second and more important is the time dependence of the (dual) Lie algebra element $t_3$. To exhibit the consequences, we compute the action for the left and right group elements\footnote{In this calculation, for $2 \times 2$ Pauli matrix generators, the definition of the trace contains a  factor of $-i$.}:
\begin{equation}
S = \frac{1}{2} \text{Tr} (t_3 \sigma_3 g_{l}^{-1} \partial_\tau g_l)+\frac{1}{2} \text{Tr} (t_3 \sigma_3  \partial_\tau g_r g_{r}^{-1})
= t_3 \dot{\psi} + \frac{t_3}{2} (\dot{\phi} \cos \theta + \dot\phi_r \cos\theta_r) \, .
\end{equation}
The extra feature (compared to e.g. \cite{Alekseev:1988vx}) is due to the term proportional to $\psi$ which in the presence of a time-dependent $t_3$ is significant. 

We first perform the path integration on $\psi$. Due to the fact that $\psi$ is periodic and because we take a periodic world line, we have a sum over winding configurations. (See e.g. \cite{Alekseev:1988vx} for similar details.) That sum discretizes the coefficient $t_3$ to take integer values.
The path integral over $\psi$ inserts a delta-function:
\begin{equation}
\text{PI}(\psi) = \sum_{n \in \mathbb{N}_0} \delta_{t_3,n} \, .
\end{equation}
We plug this into the full path integral to find:
\begin{align}
Z(\cdots)
&= \sum_{n \in \mathbb{N}_0}  
 \int (\frac{n}{2}\sin \theta   \,  d\phi \wedge d \theta )
e^{i \frac{n}{2} \int d \tau  \dot{\phi} \cos \theta}
 \int  (\frac{n}{2} \sin\theta_r d \phi_r \wedge d \theta_r )
 e^{i\frac{n}{2} \int d \tau \dot{\phi_r} \cos \theta_r} (\cdots)
 \nonumber
\, .
\end{align}
We dropped the $n=0$ term since it is zero and we used the $\mathbb{Z}_2$ Weyl symmetry to render $n$ positive. 
After performing the co-adjoint orbit quantization with Hamiltonian insertions (\ref{HamiltonianInsertions}), we find the result of the path integral \cite{Alekseev:1988vx,Alekseev:1990mp}:
\begin{align}
Z(h_l,h_r) &=\sum_{j \in \frac{1}{2} \mathbb{N}} \chi_{j} (h_l) \chi_{j} (h_r) \, ,
\end{align}
where $n=2j+1$. 
Thus, we briefly reviewed and illustrated the general procedure in \cite{Alekseev:1990mp}. We wish to argue that a similar logic holds for a large class of non-compact groups.

\subsection{A Particle on a Non-Compact Group}
In this subsection, we again follow the logic outlined in \cite{Alekseev:1990mp}, but apply it to the new domain of non-compact groups. Important differences arise. In this paper, we concentrate on the simple example of  $G=SL(2,\mathbb{R})$ that illustrates some of these differences. We will see that the integration over Lie algebra orbits splits into non-trivially distinct cases -- a difference that follows from the fact that the Killing metric on the Lie algebra has an indefinite signature.

\subsubsection{The Orbit Integrals}
 Before we delve into the details, we describe our expectations. We again calculate the path integral in steps. 
We conjugate the (dual) Lie algebra element $t$ which takes values in a Minkowski space $\mathfrak{sl}(2,\mathbb{R})=\mathbb{R}^{2,1}$ into a Cartan subalgebra. We concentrate on the generic cases in which the  Lie algebra element $t$ is future time-like, past time-like or space-like.
These cases will also be referred to as discrete plus, discrete minus and continuous. (See e.g. \cite{Ashok:2022thd} for details.) The integral over the  Lie algebra splits into a sum of three integrals accordingly.
The integration over the left and right group elements will again build coupled  left and right representation spaces.
We know that the future and past time-like Lie algebra elements give rise to discrete plus and discrete minus representations under co-adjoint orbit quantization, and the space-like Lie algebra elements to continuous representations \cite{Vergne}. (See \cite{Ashok:2022thd} for details.) We therefore expect a path integral result for periodic maps and with left and right Hamiltonian insertions:
\begin{equation}
Z(h_l, h_r) = \sum_{\stackrel{\alpha =\pm} {j =1,\frac{3}{2}\dots}  }  \chi_{j,\alpha}(h_l)
\chi_{j,\alpha}(h_r) + \sum_{\epsilon =0,\frac{1}{2} } \int_c ds \rho(s) 
\chi_{s,\epsilon}(h_l)
\chi_{s,\epsilon}(h_r)\, ,
\end{equation}
where $\epsilon=0,1/2$ corresponds to representations that represent minus the identity in SL$(2,\mathbb{R})$ trivially or non-trivially. 
We aim to understand in a little more detail the summation over discrete representations and the integral over continuous representations from the  path integral.
To make our discussion efficient, we stress only the differences with the case of the group $G=SU(2)$, and for the co-adjoint orbit quantization we lean heavily on the results derived in \cite{Ashok:2022thd}.

\subsubsection{The Differing Details}
The most convenient parameterization of the group element $g_l$ depends on whether the (dual) Lie algebra element $t_d$ is time-like (elliptic) or space-like (hyperbolic).  When the element is time-like, we choose the parameterization:
\begin{equation}
g_l =\left( \begin{array}{cc}
\cos t \cosh \rho + \cos \phi \sinh \rho  &  \sin t \cosh \rho - \sin \phi \sinh \rho \\
-\sin t \cosh \rho - \sin \phi \sinh \rho & \cos t \cosh \rho - \cos \phi \sinh \rho 
\end{array} \right) \, .
\end{equation}
We put
\be
t_d=t_{\text{ell}}=t_0 \sigma_0
\ee
where $\sigma_0= i \sigma_2$ is proportional to a Pauli matrix. In this case we find
the Lie algebra measure
\be
d t = 
2 t_0 dt_0 \wedge t_0    \, 
\sinh 2 \rho \, d (\phi+t) \wedge d \rho
\, .
\ee
The treatment of the path integral over these types of Lie algebra elements goes much as in the $SU(2)$ case. The Lie algebra element $t_0$ becomes discretized upon path integration over the coordinate $\phi-t$, which appears in the $dg_r$ measure. The  integration over the other two coordinates $(\phi+t,\rho)$ corresponds to a co-adjoint orbit quantization for discrete plus and discrete minus orbits. The details can be found in \cite{Ashok:2022thd} and references therein.
This solves the discrete part of the path integral problem. 

We still need to perform the part of the path integral where the Lie algebra element $t$ diagonalizes to a hyperbolic element:
\begin{equation}
t_d = t_{\text{hyp}} = t_3 \sigma_3 \, .
\end{equation}
Because the Lie algebra element is hyperbolic, it is convenient to use a group element parameterization that includes a hyperbolic factor. On the other hand, as discussed in detail in \cite{Ashok:2022thd}, it is most direct in the co-adjoint orbit quantization procedure to parameterize the orbit in terms of an elliptic group factor. Thus, for instance, the group decomposition  SL$(2,\mathbb{R})=G=KNA$ where $K$ is elliptic, $N$ is parabolic and $A$ is hyperbolic -- an Iwasawa decomposition -- is convenient in the detailed calculations. Thus, for this case, we parameterize the group element $g_l$ as: 
\begin{equation}
g_l = \left( \begin{array}{cc}
       \cos \theta & -\sin \theta \\
       \sin \theta & \cos \theta
            \end{array} \right) 
            \left( \begin{array}{cc}
       1 & x \\
       0 & 1
            \end{array} \right) 
            \left( \begin{array}{cc}
       e^{\rho} & 0 \\
       0 & e^{-\rho}
            \end{array} \right) \, .
\end{equation}
For this parameterization, we find the action:
\begin{equation}
S_l=\frac{1}{2} \text{Tr} (t_3 \sigma_3 g_l^{-1} \partial_\tau g_l)
= t_3 \partial_\tau \rho - t_3 x \dot{\theta} \, .
\end{equation}
% Then we have a path integration over the field $\rho$:
% \begin{equation}
% \text{PI}(\rho) = \int d \rho(\tau) e^{i t_3 \dot{\rho}}
% \, .
% \end{equation}
Because $\rho$ is non-compact, the path integration over the field $\rho$ does not lead to a discretization of $t_3$. We integrate $t_3$ over a constant real line.\footnote{There is a Weyl gauge equivalence that renders $t_3$ and $-t_3$ equivalent.}
We conclude that in the case of the group $G=SL(2,\mathbb{R})$, the quantization of the orbit label $t_d$
only occurs for the discrete representations.

\subsubsection{Result}
Thus, we have understood the reason why the path integral splits, as well as the manner in which the discrete representations are summed over and the continuous representations are integrated over. This was our main objective. The rest of the derivation proceeds exactly along the lines in \cite{Alekseev:1988vx,Alekseev:1990mp}, but now using the results for the co-adjoint orbit quantization in  \cite{Ashok:2022thd}. We record only the final result:  
\begin{align}
Z(h_l,h_r) &= \sum_{j=1,\frac{3}{2},\dots} (\chi_{j}^+(h_l) \chi_{j}^+(h_r)
+\chi_{j}^-(h_l) \chi_{j}^-(h_r))
 + \int \frac{ds}{2 \pi}  \sum_{\epsilon=0,\frac{1}{2}} \chi_{s,\epsilon}^c (h_l)
\chi_{s,\epsilon}^c(h_r) \, . \label{TwistedPartition}
\end{align}
This is in accord with the standard Hilbert space decomposition of the space ${\cal L}^2(G)$ of quadratically integrable functions on the group. 

For a more general non-compact group $G$, we expect that the final sum and integration are over the  representations in the 
 reduced dual $\hat{G}_r$ to the non-compact group  $G$, namely the set of representations occurring in the decomposition of the space ${\cal L}^2(G)$ of quadratically integrable functions \cite{Vergne}. Extending our reasoning to covers of the group SL$(2,\mathbb{R})$ is possible along the lines of \cite{Ashok:2022thd}. We will use the result in the following. It would certainly be interesting to determine more precisely the class of non-compact groups to which this path integral treatment of particle models can be generalized -- our focus in this paper lies elsewhere.  

\section{A Particle  on the BTZ Orbifold}
\label{ParticleOnOrbifold}
\label{Particle}
We obtained the partition function of a particle on the non-compact group manifold $\widetilde{G}$ twisted by global symmetry group elements. In this section, we exploit the twisted partition function to obtain the partition function for a particle on the $\mathbb{Z}$ orbifold of $\widetilde{G}=AdS_3$ which is the geometry of the BTZ black hole. The generator of the BTZ $\mathbb{Z}$ orbifold is hyperbolic \cite{Banados:1992gq}. We compare the BTZ orbifold to the case of a $\mathbb{Z}$ orbifold of $AdS_3$ which is elliptic, namely the compactification of the time direction. The former is a Lorentzian counterpart to the Euclidean black hole while the latter is the counterpart to finite temperature Euclidean $AdS_3$. We take the occasion to discuss to what extent  Lorentzian partition sums  can be related by direct analytic continuation to their Euclidean cousins.

\subsection{The Space-time Interpretation}
Our particle model will serve in part as a toy model for string theory. To that end it is useful to 
think of the particle theory with a quadratic Casimir Hamiltonian   as a first quantized model arising from the introduction of an einbein on the world line \cite{Polchinski:1998rq}. This conceptual modification also allows us to compare the first quantized results to second quantized field theory results in $AdS_3$ or the black hole background. This modification entails two technical changes to the set-up. We introduce a mass $m$ for our particle excitation as well as the constraint on the quadratic Casimir:
\begin{equation}
-j(j-1) + \frac{l^2 m^2}{4} = 0 \label{ParticleOnShell} \, ,
\end{equation}
arising from the equation of motion of the einbein. Here, the curvature scale of the $AdS_3$ space-time is denoted $l$. Secondly, even after gauge fixing the einbein, we need to integrate over a modulus with the correctly normalized measure \cite{Polchinski:1998rq}. Note that for positive mass squared, there are no continuous representation solutions $j \in 1/2 + i \, \mathbb{R}_0$ to the on-shell constraint (\ref{ParticleOnShell}).\footnote{More precisely, only discrete modes will  satisfy the Breitenlohner-Freedman bound.} The discrete plus $D^+_j$ modes are standard particle excitations in $AdS_3=\widetilde{G}$ and the discrete minus modes are their negative energy counterparts.
The same on shell equation (\ref{ParticleOnShell}) holds for the BTZ orbifold, and the same conclusions, regarding the Casimir of on-shell modes, continue to hold.

\subsection{The Euclidean Partition Function}
In Appendix \ref{AdS3} we take  the twisted Lorentzian partition function
 for a particle on $AdS_3$ as a starting point and perform an elliptic $\mathbb{Z}$ orbifold in order to compactify the time direction. We then compare the resulting Lorentzian partition function (with compactified time direction) to the one for a particle in thermal $AdS_3$ which is the Euclidean continuation of the geometry. The partition sums agree. This is no small feat since the on-shell $AdS_3$ modes are in the discrete representations while the unitary representations of the isometry group of the Euclidean model are necessarily continuous.  Our goal in the rest of the section is to perform an analogous comparison exercise for the hyperbolic BTZ orbifold. The hyperbolic orbifold is a qualitatively different Lorentzian continuation of an equivalent Euclidean geometry.

To relate the Lorentzian hyperbolic orbifold to the  Euclidean BTZ partition function for a point particle, we start by reviewing the latter. One way to obtain the partition function is as a relabelling of the thermal $AdS_3$ partition function. The one-loop result is \cite{Giombi:2008vd}:
\begin{align}
Z_{BTZ}^{particle} &= \sum_{n=1}^\infty 2 \sqrt{\pi k_b} r_+ \int_0^\infty \frac{d t}{t^{3/2}}
 \frac{e^{-\frac{k_b n^2}{t} (2 \pi r_+)^2 } e^{-\frac{t}{4k_b}(m^2+1) }}{(4 |\sin n \pi (r_--ir_+)|^{2})} \label{Projection}
 \end{align}
 where we introduced the alternative notation  $k_b=l^2/\alpha'$ for the cosmological constant scale which naturally arises from the string model to be discussed in the next section. We can expand the denominator using that $r_+>0$:
 \begin{align}
Z_{BTZ}^{particle}&=\sum_{n=1}^\infty 2 \sqrt{\pi k_b} r_+ \int_0^\infty \frac{d t}{t^{3/2}}
e^{-\frac{k_b n^2}{t} (2 \pi r_+)^2 } e^{-\frac{t}{4k_b}(m^2+1) } \sum_{r,\bar{r}=0}^\infty
e^{-2 \pi i n (r_--i r_+) (r+\frac12)} e^{ 2 \pi i n (r_-+i r_+) (\bar{r}+\frac12)} \, . \nonumber
\end{align}
We can moreover borrow a technique from the analysis of thermal string spectrum in $AdS_3$ \cite{Maldacena:2000kv} to exhibit an integral over a continuous momentum:
\begin{align}
Z_{BTZ}^{particle}&=  \sum_{n=1}^\infty \int_0^\infty  \frac{d t}{k_b \pi i n} \int_{-\infty}^\infty ds\, s\, e^{ 4 \pi i r_+ n s-\frac{s^2 t}{k_b}  -\frac{t}{4k_b}(m^2+1) } \sum_{r,\bar{r}=0}^\infty
e^{-2 \pi i n (r_--i r_+) (r+\frac12)} e^{ 2 \pi i n (r_-+i r_+) (\bar{r}+\frac12)}
\nonumber \\
&= \frac{1}{ \pi i  k_b }
\sum_{n =1}^\infty \frac{1}{n}
\int_{-\infty}^{+\infty} d s \, s  \frac{k_b}{s^2+\frac{m^2+1}{4}}
e^{4  \pi i r_+ n s}  \sum_{r,\bar{r}=0}^\infty
e^{-2 \pi i n (r_--i r_+) (r+\frac12)} e^{ 2 \pi i n (r_-+i r_+) (\bar{r}+\frac12)}~. \nonumber
\end{align}
If we close the contour in the upper half plane and pick up the pole at $j=\frac12+is=\frac12+\frac12\sqrt{m^2+1}$ we find:
\begin{align}
Z_{BTZ}^{particle} &= \phantom{-}
\sum_{n =1}^\infty \frac{1}{n}
e^{-2  \pi  r_+ n \sqrt{m^2+1}} \sum_{r,\bar{r}=0}^\infty
e^{-2 \pi i n (r_--i r_+) (r+\frac12)} e^{ 2 \pi i n (r_-+i r_+) (\bar{r}+\frac12)} \label{Logarithm} \\
 &= - \sum_{r,\bar{r}=0}^\infty \log (1-e^{-2  \pi  r_+  (2j-1)} 
e^{-2 \pi i (r_--i r_+) (r+\frac12)} e^{ 2 \pi i  (r_-+i r_+) (\bar{r}+\frac12)})\, . \nonumber
 \end{align}
 It is well-known that this partition function is a $\mathbb{Z}$ orbifold of the Euclidean $H_3$ model \cite{Giombi:2008vd}. Moreover, the Euclidean one loop determinant can be rewritten as the partition function of a second quantized scalar field:
\begin{equation}
e^{Z_{BTZ}^{particle}} = \prod_{r,\bar{r}=0,1,\dots} \frac{1}{(1-e^{-2 \pi i (r_--ir_+)(j+r)} e^{2 \pi i (r_-+ir_+)(j+\bar{r})})}
\label{SecondQuantized} \, .
\end{equation}
Consider the full exponent of a given single particle excitation. It equals:
\begin{equation}
\text{single particle exponent} = -2 \pi  (r_++ir_-) (j+r) -2 \pi  (r_+-i r_-) (j+\bar{r}) = -2 \pi i L \, , \label{ParticleExponent}
\end{equation}
where $j+r$ and $j+\bar{r}$ are eigenvalues of left/right-moving (elliptic in the Euclidean, hyperbolic in the Lorentzian) group generators. The total exponent is nothing but $-2 \pi i$ times the particle angular momentum $L$. See e.g. \cite{Ashok:2021ffx} where this exponent was discussed in detail for the case of a fundamental string in the black hole background.
In the one-loop particle amplitude the exponent $n$ of the projection operator (in equation (\ref{Projection})) also  plays the role of a multi-particle excitation number (e.g. in equation (\ref{Logarithm})) which allows us to exponentiate the first quantized formula into a neat second quantized quantum field theory result (\ref{SecondQuantized}). 

Remarkably, the same formula (\ref{SecondQuantized}), after transformation, 
can be interpreted as the spectrum of single particle quasinormal modes on the black hole background with in-going boundary conditions at the horizon and a no-energy-loss requirement at infinity \cite{Denef:2009kn}.  We have the equality \cite{Denef:2009kn,Keeler:2018lza}:
\begin{equation}
Z^{particle}_{BTZ} = - \sum_{n>0,p\ge 0}(\log (1-q^{n+p+j} \bar{q}^{p+j})+ \log(1-q^{p+j} \bar{q}^{n+p+j})) -\sum_{p \ge 0} \log(1-(q \bar{q})^{p+j}) \, 
\end{equation}
where $q=e^{-2 \pi i (r_--ir_+)}$. The exponents can be identified as the single particle quasinormal mode spectrum \cite{Birmingham:2001pj} when $j$ is tuned such that the quasinormal mode energy coincides with a Matsubara frequency $\omega_{QN}(j)=\omega_{Matsubara}$ \cite{Denef:2009kn}. In other words, the poles of the partition function code the quasinormal mode frequencies and vice versa. 
The resulting spectrum coincides with the one obtained by direct calculation \cite{Birmingham:2001pj}:
\begin{equation}
E = \pm L -i (r_+ \mp r_-)(2j+2p) \, .
\label{ParticleQNM}
\end{equation}
We have thus reviewed that there is both a Euclidean orbifold reading of the partition function in terms of a projection operator {\em and} an interpretation in terms of a quasinormal spectrum.

\subsection{The BTZ Lorentzian Hyperbolic Orbifold}
We wish to circle back now and interpret the Euclidean partition function result in terms of the Lorentzian representation theory we introduced previously. 
To that end, we note the intermediate result -- see equation (\ref{Logarithm}) --:
\begin{equation}
Z_{BTZ}^{particle} = 
\sum_{n =1}^\infty \frac{1}{n}
e^{-2  \pi  r_+ n (2j-1)} \sum_{p,\bar{p}=0}^\infty
e^{-2 \pi i n (r_--i r_+) (p+\frac12)} e^{ 2 \pi i n (r_-+i r_+) (\bar{p}+\frac12)} \, .
\end{equation}
Recall that  the  discrete character of a hyperbolic group element is \cite{Ashok:2022thd}:
\begin{align}
\chi_j^+(e^{t \, t_{\text{hyp}}}) 
%= \Tr\, T_{j}^+ (e^{t \, t_{\text{hyp}}}) &=
%e^{\pm 2 \pi i p j}
&=\frac{e^{-|2j-1||t|}}{|\sinh t|} \, .
\end{align}
We can also rewrite the  result as 
\begin{equation}
Z_{BTZ}^{particle} =
\sum_{n > 0} 
%\frac{e^{-4 \pi r_+ n \sqrt{m^2+1}/2}}{n}
\frac{1}{n} \chi_{j}^+(e^{\pi n (r_++ir_-)t_{\text{hyp}}}) \chi_{j}^+(e^{\pi n (r_+-ir_-)t_{\text{hyp}}}) \, .
\end{equation}
We sketch how this result arises from the  direct calculation of the Lorentzian BTZ orbifold partition function.\footnote{In doing so, we will assume that $ir_-$ is real which is true after analytic continuation to the Lorentzian.}
We obtain the partition function for a particle on the BTZ orbifold from the twisted partition function on the universal cover $\widetilde{G}$. The twisted partition function on the cover is:
\begin{align}
Z(h_l,h_r) &=  \int_{1/2}^{\infty} {dj} (\chi_{j}^+(h_l) \chi_{j}^+(h_r)
+\chi_{j}^-(h_l) \chi_{j}^-(h_r))
%\nonumber \\
%& 
+ \int_0^\infty ds  \int_0^1 d \epsilon \chi_{s,\epsilon}^c (h_l)
\chi_{s,\epsilon}^c(h_r) \, .
\end{align}
We wish to insert an orbifold projection operator in the trace which projects onto states that are invariant under the hyperbolic $\mathbb{Z}$ orbifold. To that end, we sum over the insertion of a group generator $(h_l,h_r)=(e^{\pi (r_++ir_-)t_{\text{hyp}}},e^{\pi(r_+-ir_-)t_{\text{hyp}}})$ raised to an arbitrary power $n \in \mathbb{Z}$:
\begin{align}
Z^{particle}_{BTZ} =&\sum_{n \in \mathbb{Z}} \int_{1/2}^{\infty} {dj} \, \, \Big(\chi_{j}^+(e^{\pi n (r_++ir_-)t_{\text{hyp}}}) \chi_{j}^+(e^{\pi n (r_+-ir_-)t_{\text{hyp}}}) \nonumber \\
& \hspace{1.8cm}
+\chi_{j}^-(e^{\pi n (r_++ir_-)t_{\text{hyp}}}) \chi_{j}^-(e^{\pi n (r_+-ir_-)t_{\text{hyp}}}) \Big)
\nonumber \\
& + \int_0^\infty ds  \int_0^1\, d \epsilon~ \chi_{s,\epsilon}^c (e^{\pi n (r_++ir_-)t_{\text{hyp}}})
\chi_{s,\epsilon}^c(e^{\pi n (r_+-ir_-)t_{\text{hyp}}})  \, . \label{OrbifoldBTZ}
\end{align}
The physical state condition (\ref{ParticleOnShell}) projects this partition function onto the spin $j$ associated to the mass $m$ of the state under consideration. In the calculation of the Euclidean partition function, it was the projection of the contour integral onto the pole that played the role of the on-shell projection. This is implemented by the Schwinger and momentum integral.  
 The prefactor $1/n$ arises from these integrals in the same manner as in the Euclidean partition function calculation. 
By comparing the Lorentzian orbifold partition function (\ref{OrbifoldBTZ}) with the Euclidean partition function calculation, we again recognize that the multi-particle number $n$ doubles as the order of the orbifold projection operator in the partition function (\ref{OrbifoldBTZ}).  It is important in making this match that the discrete character is a function of the smallest hyperbolic eigenvalue, thus rendering the character even in $n$.\footnote{The background independent term  $n=0$ can again be ignored.}

\subsubsection{Summary}
We conclude that the Euclidean particle partition function on the  BTZ black hole affords a reading as an on-shell hyperbolic Lorentzian orbifold with respect to the angular momentum operator $L$. Simultaneously, it can be read as the result of tracing over  the black hole quasinormal modes. In the next section, we generalize this treatment to the case of the Euclidean fundamental string partition function.

\section{The String Spectrum on a Black Hole}
\label{Strings}
In this section, we revisit the fundamental string spectrum on the BTZ black hole background with NSNS flux \cite{Natsuume:1996ij,Hemming:2001we,Troost:2002wk,Hemming:2002kd,Rangamani:2007fz,Berkooz:2007fe,Mertens:2015ola,Ashok:2021ffx,Nippanikar:2021skr}. We improve on the analysis of the world sheet partition function \cite{Rangamani:2007fz,Berkooz:2007fe,Mertens:2015ola,Ashok:2021ffx,Nippanikar:2021skr} and acquire a tentative but conjecturally rather complete picture of its spectral content. All along the analysis it is important to keep the lessons in mind that we take away from the particle partition functions.

The  interpretation of the string partition function poses considerable extra hurdles. The boundary circles of the Euclidean geometry allow for winding modes and we need to disentangle the associated novelties. Because the thermal circle in the $AdS_3$ background is topologically non-trivial while in the black hole background it is topologically trivial, the analytic continuation to the Lorentzian proceeds quite differently in regard to the winding modes.  Thus, while our analysis certainly draws upon the knowledge acquired in the context of thermal $AdS_3$ \cite{Maldacena:2000kv}, it follows  a very different logic and this is reflected in the course of the calculation. 

\subsection{The BTZ Partition Function}
Starting from the one loop string amplitude for thermal $AdS_3$ \cite{Maldacena:2000kv},
one obtains the corresponding result for the Euclidean BTZ black hole: 
\begin{align}
\label{ZandZBTZsaddle}
Z &= \int_F \frac{d^2 \tau}{2 \tau_2} Z_{BTZ} Z_{gh} Z_{int}
 \\
 Z_{BTZ}&= \frac{2r_+ \sqrt{k_b-2}}{ \sqrt{\tau_2}}
\sum_{m,w} \frac{e^{- \pi \frac{k_b}{\tau_2} r_+^2 |m-w \tau|^2+ \frac{2 \pi}{\tau_2} \text{Im}(\bar{U}_{m,w})^2}}{|\theta_1(\bar{U}_{m,w},\tau)|^2}  \label{ZBTZ}
\, .
\end{align}
We have defined ghost non-zero modes and internal partition function factors $Z_{gh} = |\eta(\tau)|^4$ and $Z_{int}$. The internal conformal field theory is a compact conformal field theory with central charge $c_{int} = 26 - \left( 3 + \frac{6}{k_b-2} \right)$ where $k_b$ is the level of the $\widetilde{G}$ Wess-Zumino-Witten model. We have also denoted the holonomy on which the path integral depends as 
\be
\bar U_{m,w} = (r_- - i r_+)(m-w\tau)
\, .
\ee
For the notation and further details we refer to \cite{Ashok:2021ffx}.
In brief, the formula is obtained by mapping the inverse temperature $\beta$ and the fugacity $\beta \mu$ for the angular momentum on the thermal $AdS_3$ side \cite{Maldacena:2000kv} to the outer and inner horizon radius of the  black hole \cite{Carlip:1994gc,Maldacena:1998bw}:
\be 
(\frac{\beta}{2\pi}, \frac{\beta \mu}{2\pi}) \leftrightarrow (r_+, r_-)~.  
\ee 
The path integral in the Euclidean geometries are identical because the geometries are related by an S-duality transformation on the boundary torus \cite{Maldacena:1998bw}.

One open question about this partition function is whether one can rewrite it as a sum of logarithms which is necessary to find an interpretation of the spectrum in terms of  single particle states. In this section we give a detailed  answer to this question in two parts: first of all we exploit the  methods of \cite{Maldacena:2000kv}  and perform a saddle point approximation to the partition function. It is the discrete representations of sl$(2,\mathbb{R})$ that contribute to the saddle points and we find a stringy generalization of the particle calculation performed in the previous section, leading to an interpretation of this part of the spectrum in terms of  quasinormal winding modes. We also provide a space-time description of these modes.
In a second part,  we analyze the partition function near the poles of the integrand and interpret the contribution near the poles as arising from  multiply wound long strings. This contribution lies in  the irreducible continuous representations of sl$(2,\mathbb{R})$ and it has a distinct dependence on the boundary modular parameter.

\subsubsection{Affine Character Reading}
As a preface to expanding and contracting the formula (\ref{ZBTZ}) for the partition function, we attempt to read it at face value. We identify the one loop vacuum amplitude as a sum over images of a product of hyperbolic affine characters. 
To that end we conjecture the existence of hyperbolic affine characters -- they are generalizations of sl$(2,\mathbb{R})$ hyperbolic characters --:
\begin{equation}
\hat{\chi}(\tau,t) =  \frac{\cos 2ts}{|\sinh t|} \frac{1}{\prod_n(1 -e^{2t} q^n)(1-e^{-2t} q^n) (1-q^n)} \, .
\label{AffineContinuousCharacter}
\end{equation}
In Appendix \ref{HyperbolicAffineCharacters} we provide first arguments for why these proposed characters are sensible, based on the  decomposition of the tensor product of infinite and finite dimensional representations of sl$(2,\mathbb{R})$. 
Given this character, we identify its denominator as proportional to the  $\theta_1$ factor in the partition function where we make the identification $i \pi \bar{U}_{m,w}=t$.
As in the particle case, we surmise that we are  summing over orbifold images labelled by $m$. Moreover, as always in conformal field theory orbifolds, we have   introduced twisted sectors labelled by the winding number $w$. There are however important issues that remain to be understood: first of all, we need to  identify the correct spectrum of momenta $s$ that is integrated over in the partition function and whether there are  discrete representation characters that  contribute in the Lorentzian theory; secondly we need to determine which part of the world sheet spectrum contributes on-shell in space-time. We  answer these questions in the next two subsections.

\subsection{The Saddle Points and the Quasinormal Winding Modes}
In this subsection, we mould the partition function into a form in which we can perform a telling saddle point analysis. It will allow us to identify the contribution of quasinormal winding modes to the one-loop amplitude. 

We first introduce the radial momentum $s$-integral as in \cite{Ashok:2021ffx}:
\begin{align}
\int_{-\infty}^{+\infty} ds~ e^{-4 \pi \tau_2 \frac{s^2}{k_b-2}} e^{-4 \pi i s \text{Im} (\bar{U}_{m,w})} &= \frac{1}{2} \frac{\sqrt{k_b-2}}{\sqrt{\tau_2}}
e^{ -\frac{\pi (k_b-2)}{\tau_2}  \text{Im}(\bar{U}_{m,w})^2}
\, .
\end{align}
This allows one to rewrite the BTZ factor of the one loop integrand as:
\begin{align}
Z_{BTZ} &= 4r_+ 
\sum_{m,w} 
\int_{-\infty}^{+\infty} ds e^{-4 \pi \tau_2 \frac{s^2}{k_b-2}} e^{-4 \pi i s Im (\bar{U}_{m,w})}
\frac{e^{- \pi \frac{k_b}{\tau_2} r_+^2 |m-w \tau|^2+ \frac{k_b \pi}{\tau_2} Im(\bar{U}_{m,w})^2}}{|\theta_1(\bar{U}_{m,w},\tau)|^2} 
\, .
\end{align}
We  follow the route taken in \cite{Ashok:2020dnc} that makes it possible to expand the $\theta$-function in the denominator. For this purpose we introduce the delta-function:
\begin{align}
1 & = \tau_2 \int_0^1 d^2 s \sum_{v,w'} \delta^2(\bar U_{m,w}-(s_1+w')\tau+s_2+v) \, .
\end{align}
We now implement a trick, inspired by (but not identical to) the unfolding of thermal partition functions in string theory \cite{Polchinski:1985zf}, which uses the modular invariance of all factors in the integrand of the integral over the fundamental domain. Given the invariance, each factor is identical in each copy of the fundamental domain. Moreover, under modular transformations, the pair of integers $(v,w')$ transforms as a doublet. For this particular delta-function factor, instead of integrating a modular invariant expression over a single fundamental domain, we integrate the vectors $(v,0)$ over all copies of the
fundamental domain in the strip. This restores the full doublet when transformed back to the fundamental domain. These integrations are therefore identical and because all other factors in the integrand are modular invariant, they go along for the ride.  Thus, we set
 $w'=0$ from now on, departing strongly from the analysis of thermal partition functions in $AdS_3$, and unfold the Schwinger $\tau_2$-integral, such that we  integrate over the strip in the $\tau$-plane, with $|\tau_1| < \frac12$ and $0 < \tau_2 < \infty$. 
We follow this up by introducing the integral representation for the $\delta$ function \cite{Ashok:2020dnc}:
\begin{align}
\delta^2 (\bar U_{m,w}-s_1\tau+s_2+v) &=
\int d^2 \lambda_i  e^{2 \pi i \lambda_1 (r_-(m-w\tau_1) - r_+ w \tau_2 -s_1 \tau_1 + s_2 + v)}
e^{2 \pi i \lambda_2 (-r_+(m-w\tau_1) -r_-w\tau_2 -s_1 \tau_2) }\, .
\end{align}
By making use of the elliptic properties of the $\theta$-function and after simplification we find that 
% \begin{align}
% Z_{BTZ} &=4 r_+  \tau_2 \int_0^1 d^2 s \sum_{v}
% \sum_{m,w} \int d^2 \lambda_i
% \int_{-\infty}^{+\infty} ds e^{-4 \pi \tau_2 \frac{s^2}{k_b-2}} e^{-4 \pi i s \text{Im} (\bar{U}_{m,w})}
% \frac{e^{- \pi \frac{k_b}{\tau_2} r_+^2 |m-w \tau|^2+ \frac{k_b \pi}{\tau_2} Im(\bar{U}_{m,w})^2}}{|\theta_1(s_1 \tau-s_2,\tau)|^2} \nonumber \\
% &\hspace{3cm} e^{2 \pi i \lambda_1 (r_-(m-w\tau_1) - r_+ w \tau_2 -s_1 \tau_1 + s_2 + v)}
% e^{2 \pi i \lambda_2 (-r_+(m-w\tau_1) -r_-w\tau_2 -s_1 \tau_2) }
% \, .
% \end{align}
% % We should immediately simplify the initial combination of exponents:
% \begin{align}
% - \pi \frac{k_b}{\tau_2} r_+^2 |n-m \tau|^2+ \frac{k_b \pi}{\tau_2} Im(\bar{U}_{n,m})^2 &= 2\pi k_b m  n r_+ r_- + \pi k_b m^2 (-2 r_+ r_- \tau_1 + r_-^2 \tau_2-r_+^2 \tau_2)
% \end{align}
%Simplifying the exponent, we have the unfolded integrand: 
\begin{align}
Z_{BTZ} &= 4r_+  \tau_2 \int_0^1 d^2 s 
\sum_{m,w,v} \int d^2 \lambda_i
\int_{-\infty}^{+\infty} ds \frac{e^{-4 \pi \tau_2 \frac{s^2}{k_b-2}} e^{-4 \pi i s \text{Im} (\bar{U}_{m,w})}}
{|\theta_1(s_1 \tau-s_2,\tau)|^2}\, e^{2 \pi i \lambda_2 (-r_+(m-w\tau_1) -r_-w\tau_2 -s_1 \tau_2)}
 \nonumber \\
&\hspace{2cm}
e^{2\pi k_b m  w r_+ r_- + \pi k_b w^2 (-2 r_+ r_- \tau_1 + r_-^2 \tau_2-r_+^2 \tau_2)}
e^{2 \pi i \lambda_1 (r_-(m-w\tau_1) - r_+ w \tau_2 -s_1 \tau_1 + s_2 + v)}
\, .
\end{align}
Given the range of the holonomy $s_1$ one can   expand the $\theta$-functions: 
\be 
\frac{1}{|\theta_1(\nu, \tau)|^2} = \frac{ \sqrt{z \bar z}}{|\eta(\tau)|^6}\sum_{r,\bar r \in \mathbb{Z}} z^r~ \bar z^{\bar r}~ S_r S_{\bar r}~,   
\ee 
where $z = e^{2\pi i \nu}$ is the fugacity and the series $S_r$ is given by  $S_r = \sum_{n=0}^{\infty} (-1)^n q^{\frac{n}{2}(n+2r+1)}$. 
We   list further steps performed in detail in \cite{Ashok:2020dnc}: i) We perform the sum over $v$, which leads to a Dirac comb for $\lambda_1$. ii) The integral over $s_2$ leads to the constraint $\lambda_1 = r-\bar r \in \mathbb{Z}$. iii) These two combine and the result is a  trivial integral over $\lambda_1$. Taking these three steps we obtain
\begin{align}
Z_{BTZ} 
%&= r_+  \tau_2 \int_0^1 d^2 s 
% \sum_{n,m,l_1} \int d \lambda_2
% \int_{-\infty}^{+\infty} ds e^{-4 \pi \tau_2 \frac{s^2}{k_b-2}} e^{-4 \pi i sIm (\bar{U}_{n,m})}
% \frac{e^{2\pi k_b m  n r_+ r_- + \pi k_b m^2 (-2 r_+ r_- \tau_1 + r_-^2 \tau_2-r_+^2 \tau_2)}}{|\theta_1(s_1 \tau-s_2,\tau)|^2} \nonumber \\
% & e^{2 \pi i l_1 (r_-(n-m\tau_1) - r_+ m \tau_2 -s_1 \tau_1 + s_2 )}
% e^{2 \pi i \lambda_2 (-r_+(n-m\tau_1) -r_-m\tau_2 -s_1 \tau_2) }
% \nonumber \\
% &=  r_+  \tau_2 \int_0^1 d^2 s 
% \sum_{n,m,l_1} \int d \lambda_2
% \int_{-\infty}^{+\infty} ds e^{-4 \pi \tau_2 \frac{s^2}{k_b-2}} e^{-4 \pi i sIm (\bar{U}_{n,m})}
% e^{2\pi k_b m  n r_+ r_- + \pi k_b m^2 (-2 r_+ r_- \tau_1 + r_-^2 \tau_2-r_+^2 \tau_2)}\nonumber \\
% & e^{2 \pi i l_1 (r_-(n-m\tau_1) - r_+ m \tau_2 -s_1 \tau_1 + s_2 )}
% e^{2 \pi i \lambda_2 (-r_+(n-m\tau_1) -r_-m\tau_2 -s_1 \tau_2) }
% \nonumber \\
% & \frac{1}{|\eta|^6} \sum_{r,\bar{r}}
% e^{-2 \pi s_1 \tau_2 (r+ \bar{r}+1)} e^{2 \pi i (s_1 \tau_1 -s_2) (r-\bar{r})}  S_r S_{\bar{r}}
% \nonumber \\
&=  4r_+  \tau_2 \int_0^1 d s_1 
\sum_{m,w} \int d \lambda_2
\int_{-\infty}^{+\infty} ds e^{-4 \pi \tau_2 \frac{s^2}{k_b-2}} e^{-4 \pi i s \text{Im} (\bar{U}_{m,w})}
e^{2\pi k_b m  w r_+ r_- + \pi k_b w^2 (-2 r_+ r_- \tau_1 + r_-^2 \tau_2-r_+^2 \tau_2)}\nonumber \\
& \sum_{r,\bar{r}} e^{2 \pi i (r-\bar{r})(r_-(m-w\tau_1) - r_+ w \tau_2  )}
e^{2 \pi i \lambda_2 (-r_+(m-w\tau_1) -r_-w\tau_2 -s_1 \tau_2) }
 \frac{1}{|\eta|^6} 
e^{-2 \pi s_1 \tau_2 (r+ \bar{r}+1)}  S_r S_{\bar{r}}
\, .
\end{align}
We   use the value of the imaginary part of the holonomy $\text{Im}\, U_{m,w} = -(r_+(m-w \tau_1) + r_-w \tau_2)$ 
% to obtain
% %
% \begin{align}
% Z &= r_+ \int \frac{d^2\tau}{2}  \int_0^1 d s_1 
% \sum_{m,w} \int d \lambda_2
% \int_{-\infty}^{+\infty} ds e^{-4 \pi \tau_2 \frac{s^2}{k_b-2}} e^{2 \pi i (2s - \lambda_2) (r_+(m-w \tau_1) + r_-w \tau_2) }\cr
% & e^{2\pi k_b m w r_+ r_- + \pi k_b w^2 (-2 r_+ r_- \tau_1 + r_-^2 \tau_2-r_+^2 \tau_2)}
% e^{-2 \pi i \lambda_2 s_1 \tau_2 }\nonumber \\
% & \sum_{r,\bar{r}} e^{2 \pi i (r-\bar{r})(r_-(m-w\tau_1) - r_+ w \tau_2  )}
%  \frac{1}{|\eta|^6} 
% e^{-2 \pi s_1 \tau_2 (r+ \bar{r}+1)}  S_r S_{\bar{r}}~Z_{gh}Z_{int}
% \, .
% \end{align}
and  the primaries plus oscillators expansion 
\be 
 \frac{1}{|\eta|^6} 
S_r S_{\bar{r}}~Z_{gh}Z_{int} = (q\bar q)^{\frac{1}{4(k-2)} -1}\sum_{h,N} d_{h,N}\, q^{h+N} \bar q^{\bar h + \bar N}~,
\ee
to further simplify the worldsheet partition function $Z_{BTZ}$. We suppress the summation over the anti-holomorphic indices and labels $(\bar r, \bar h, \bar N)$ on the degeneracies to avoid clutter. Substituting the resulting expression into the one loop string amplitude, we collect terms in $\tau_1$ and $\tau_2$ in the exponent to obtain 
\begin{align}
Z &= 4r_+ \int d^2\tau  \int_0^1 d s_1 
\sum_{m,w,r,h,N} \int d \lambda_2
\int_{-\infty}^{+\infty} ds~  d_{h,N}\, e^{2\pi i m (r_-(r-\bar r) + r_+(2s-\lambda_2) - i k_b w r_+ r_-)}\cr
& e^{2\pi i \tau_1 \big(h+N - \bar h - \bar N -w (r_- (r-\bar r) + r_+(2s-\lambda_2) - i k_b w r_+r_- ) \big) }\cr
& e^{-2\pi \tau_2\big(-2+h+N+ \bar h +\bar N + \frac{2s^2 + \frac{1}{2} }{(k_b-2) } + \frac{k_b w^2}{2}(r_+^2- r_-^2) +s_1(r+\bar r + 1 + i\lambda_2) + i w r_+(r-\bar r) - i w r_- ( 2s-\lambda_2) \big)}~.
\end{align}
We perform the $\tau_1$-integral by solving the level matching constraint for $\bar L_0$ in terms of $L_0$ first 
%
% \begin{align}
% Z &= 2r_+ \int d\tau_2 \int_0^1 d s_1 
% \sum_{m,w} \int d \lambda_2
% \int_{-\infty}^{+\infty} ds~ e^{2\pi i m (r_-(r-\bar r) + r_+(2s-\lambda_2) - i k_b w r_+ r_-)}\cr
% %& e^{2\pi i \tau_1 \big(h+N - \bar h - \bar N -w (r_- (r-\bar r) + r_+(2s-\lambda_2) - i k_b w r_+r_- ) \big) }\cr
% & e^{-2\pi \tau_2\big(-2+2h+2N + \frac{2s^2 + \frac{1}{2} }{(k_b-2) } + \frac{k_b w^2}{2}(r_++i r_-)^2 +s_1(r+\bar r + 1 + i\lambda_2) + i w (r_+ + i r_-)(r-\bar r)- (r_+ + i  r_-)w ( 2s-\lambda_2)  \big)}~.
% \end{align}
and perform the radial momentum $s$-integral next to obtain 
\begin{align}
Z &= 2r_+\sqrt{k_b-2} \int \frac{d\tau_2}{2\sqrt{\tau_2}}  \int_0^1 d s_1 
\sum_{m,w,r, h,N} \int d \lambda_2~ d_{h,N}\,
e^{2\pi i m (r_-(r-\bar r)  -r_+\lambda_2 - i k_b w r_+ r_-)}\cr
& e^{-2\pi \tau_2\big(-2+2h+2N+\frac{1}{2(k_b-2)}  + \frac{k_b w^2}{2}(r_++i r_-)^2 +s_1(r+\bar r + 1 + i\lambda_2) + i w (r_+ + i r_-)(r-\bar r)+ (r_+ + i  r_-)w \lambda_2  \big)}\cr
& e^{-\frac{\pi (k_b-2) }{\tau_2}(m r_+ + w (r_- - i r_+) \tau_2)^2 }~.
\end{align}
The $\lambda_2$-integral can  be trivially done to find the $\delta$-function constraint that allows one to perform the $s_1$-integral:
\be 
s_1\tau_2 = - m r_+ + i w \tau_2 (r_+ + i r_-)~.
\ee
At this point it is clear that if we are to find a real solution for $s_1$, we  need to perform an analytic continuation, and consider the combination $ i (r_+ + i r_-)$ to be real. Crucially, in the sequel we consider the continuation for which  $iw(r_++ir_-)$ is negative. In that case, $m$ is necessarily negative.  
Furthermore it is important to keep in mind that this constrains the range of the parameter $\tau_2$ as the solution for $s_1$ has to be in the open interval $(0,1)$.
\be 
\label{tau2bound}
0 <  -\frac{mr_+}{\tau_2}+ iw (r_+ + i r_-) < 1~. 
\ee 
Thus, the $s_1$ and $\lambda_2$-integrals can be done and  we obtain 
\begin{align}
    Z &=  2r_+\sqrt{k_b-2} \int \frac{d\tau_2}{\tau_2^{\frac{3}{2}}} \sum_{w,r,h,N} \sum_{m>0}d_{h,N}e^{2\pi i m\big(-i(r_+ + i r_-)(r+\frac12)-i(r_+ - i r_-)(\bar r +\frac12) -2 i  r_+ r_- w+ (k_b-2) r_+^2  w \big)} \cr
    &\hspace{4cm} e^{-2\pi\tau_2\left(-2+2h+2N+\frac{1}{2(k_b-2)}+iw(1+2r)(r_++ i r_-)+ w^2(r_+ + i  r_-)^2  \right)}~e^{-\frac{\pi (k_b-2) m^2 r_+^2}{\tau_2}}~,
\nonumber
\end{align}
where we have   restricted the sum to just $m<0$, and multiplied by a factor of two. After these preparations, since we have not found a method yet to continue exactly, we perform a saddle point approximation to the $\tau_2$-integral. The saddle point is at 
%
%{\scriptsize
\begin{align}
    (\tau_2)_{saddle} = - \frac{ (k_b-2) m r_+} {\sqrt{1+4(k_b-2)\big(-1 + h+N+i\frac{w}{2}(r_+ + i r_-)(1+2r) +\frac{w^2}{2}(r_+ + i r_-)^2  \big) }}~.
\end{align}
%}
We denote the argument of the square root as 
\be 
A = 1+4(k_b-2)\big(-1 + h+N+i\frac{w}{2}(r_+ + i r_-)(1+2r) +\frac{w^2}{2}(r_+ + i r_-)^2  \big) \, .
\ee We shall soon relate this argument to the spin quantum number $j$. Evaluating the partition function on the saddle point, and performing the resulting Gaussian integral, we find that  
\begin{align}
 Z_{\text{saddle}} 
 %&= \sum_{w} \sum_{m>0} \frac{\sqrt{k_b-2}\, r_+}{\tau_{saddle}^{\frac{3}{2}}}\frac{1}{2} \sqrt{\frac{(k_b-2) \tau_{saddle}}{A_{BTZ} }}  e^{2\pi i m \left( r_-(r-\bar r) - i r_+(1+r+\bar r) -2 i r_+ r_- w +  (k_b-2) r_+^2 w - i r_+\sqrt{A}  \right) } \cr
&=-\sum_{w,r,h,N} d_{h,N}\sum_{m<0} \frac{1}{m} e^{2\pi i m \left( -i(r_+ + i r_-)(r+\frac12)-i(r_+ - i r_-)(\bar r +\frac12) -2 i r_+ r_- w +  (k_b-2) r_+^2 w - i r_+\sqrt{A}  \right) } \cr
&= -\sum_{w,r, h,N}d_{h,N} \log\left(1- e^{-2\pi i  \left( -i(r_+ + i r_-)(r+\frac12)-i(r_+ - i r_-)(\bar r +\frac12) -2 i r_+ r_- w +  (k_b-2) r_+^2 w - i r_+\sqrt{A}  \right) } \right) \nonumber
\end{align}
We  have obtained a logarithm that is the result of resummation over the multiple short string contributions to the one loop amplitude.

Let us   interpret this result by exploiting properties of the worldsheet conformal field theory that describes the BTZ black hole background \cite{Natsuume:1996ij,Hemming:2001we,Troost:2002wk,Hemming:2002kd,Rangamani:2007fz,Berkooz:2007fe,Mertens:2015ola,Ashok:2021ffx,Nippanikar:2021skr}. The first details to recall are the worldsheet conformal dimensions of the short strings in the Euclidean theory. These are  \cite{Ashok:2021ffx}:
\begin{align}
    L_0 &= -\frac{j(j-1)}{k_b-2} + i(j+r)(r_+ + i r_-)w+ \frac{k_bw^2}{4}(r_+ + i r_-)^2 + h + N \nonumber \\
    \bar L_0 &= -\frac{j(j-1)}{k_b-2} - i(j+\bar r)(r_+ - i r_-)w+ \frac{k_bw^2}{4}(r_+ - i r_-)^2 + \bar h + \bar N~.  \label{L0L0barEBTZ}
\end{align}
We  imposed the level matching condition $L_0 = \bar L_0$ while integrating over $\tau_1$ (and requiring translation invariance in the world sheet coordinate $\sigma$) and the on-shell conditions therefore reduce to $L_0=1$. Solving for the spin $j$, we find the relation
\be 
2j - 1 - i (k_b-2) w (r_+ + i r_-) = \sqrt{A}~.  \label{SquareRootSpin}
\ee
This  relation  allows us to interpret the exponent that appears in  the partition function for the saddle point contribution intuitively. We note that with our analytic continuation, both the square root as well as the spin $j$ are real. Substituting for $A$, we obtain  
\begin{align}
    Z_{\text{saddle}} =- \sum_{r,h,N,w\ge 0}~d_{h,N}~ \log\left(1-e^{-2\pi i  \left( -i(r_++ i r_-)(j+r) -i(r_+ - i r_-)(j+\bar r) -i k_b r_+r_-w  \right) } \right)~.
\end{align}
We recognize the exponent as the eigenvalue of the operator $-2 \pi i L_{string}$, where $L_{string}$ is the angular momentum operator in the Euclidean theory \cite{Hemming:2001we,Ashok:2021ffx} 
\begin{align}  
L_{string} &= \frac{1}{w}(h+N - \bar h - \bar N) \cr
&=-i(r_+ + i r_-)(j+r)  - i(r_+ - i r_-) (j+\bar r) - i k_b w r_+ r_- ~.
%&= -i(r_+ + i r_-)\big(j+r -i \frac{ k_b w}{4}(r_++ i r_-)\big)  - i(r_+ - i r_-) (j+\bar r + i \frac{ k_b w}{4}(r_+ - i r_-))~.
\end{align}
Thus, the full saddle point contribution can be given the interpretation as the trace: 
\begin{equation}
Z_{\text{saddle}} = -\sum_{m<0}  \frac{1}{m}\text{Tr}_{\text{discrete saddle}} e^{2 \pi i m L_{string}} \, .
\end{equation}
The "multi-particle" number $m$ is the dual of the angular momentum (as in \cite{Ashok:2021ffx}). We recognized this as the logarithm of the single string trace:
\begin{equation}
Z_{\text{single string}} = \text{Tr}_{\text{discrete saddle}} e^{-2 \pi i  L_{string}} \, .
\end{equation}
This trace formula is a stringy generalization of the particle formula (\ref{ParticleExponent}) in section \ref{Particle}. 

\subsubsection{The Bound on the Spin}

As we have seen in the inequality \eqref{tau2bound} the constraint that $0 < s_1 < 1$ translates to a bound on $\tau_2$. Combining the saddle point equation with the expression (\ref{SquareRootSpin}) for the square root in terms of the spin one can write 
\be 
-\frac{m r_+}{(\tau_2)_{saddle}} + iw(r_+ + i r_-)= \frac{2j - 1}{k_b-2} ~. 
\ee
Substituting this into the constraint  \eqref{tau2bound} we find the bound on spin:
\begin{align}
    \frac12 & < j < \frac{k_b}{2}  ~.
\end{align}
We note that there is an intuitive interpretation of this bound. In the context of a $su(2)$ Wess-Zumino-Witten model for instance, we can argue that a spin $j$ that reaches the level $k_b/2$ will create a wiggle in space-time which is of the string scale and equivalent to an oscillator mode. Thus, the bound on the spin is interpreted as a rewriting of a higher spin excitation as a string oscillation. The same reasoning holds in sl$(2,\mathbb{R})$. When the spin $j$ reaches the curvature scale squared $k_b/2$, angular momentum turns into  string scale oscillations, also in non-compact, locally $AdS_3$ spaces. For global $AdS_3$, this bound was indeed found to hold in \cite{Maldacena:2000hw,Maldacena:2000kv}.  The physical phenomenon that we described is  independent of the global structure of the space-time and thus we may expect the bound on spin to be independent of the background parameters $(r_+,r_-)$ of the asymptotically $AdS_3$ geometry. Our saddle point analysis confirms that this is so.

\subsubsection{The Quasinormal Winding Modes}
\label{WavefunctionAnalysis}
\label{ZeroModeVertexOperators}
\label{QNWM}

We have derived through analytic continuation and saddle point evaluation of the one loop vacuum amplitude a contribution of discrete single string modes.  It is natural to ask for a more direct space-time description of this contribution. We  propose to use this contribution to identify space-time frequencies of quasinormal winding modes following the method of \cite{Denef:2009kn} which consists in exploiting poles of the partition function to determine quasinormal frequencies.
We consider the contribution to the partition function from a primary twisted string,  fixing the winding $w$ and the internal conformal dimension $h$. We concentrate on the leading terms in the descendant degeneracy function $S_r(q)$, which is a constant equal to $1$ for $r \ge 0$ (and subleading for $r<0$), and similarly for the quantum number $\bar{r}$.  In other words, we concentrate on primaries. 
Thus, we restrict to $r  \ge 0$ and $\bar{r} \ge 0$ and start out with the  expression
\begin{align}
    Z^{(w)}_{\text{saddle}}=\sum_{|m|=1}^{\infty}\sum_{r,\bar r\ge 0}\frac{1}{|m|} e^{-2\pi |m|\left( (r_+ +i r_-)(j+r-i\frac{k_b w (r_++i r_-)}{4})+ (r_+ - i r_-)(j+\bar r +i\frac{k_b w (r_+-i r_-)}{4})  \right)}
\, . \label{WSaddleContribution}
\end{align}
 The expansion of the logarithm can then be rewritten, following \cite{Denef:2009kn,Castro:2017mfj, Keeler:2018lza} as -- see also section \ref{Particle} -- :
\begin{align}
Z^{(w)}_{\text{saddle}}=& -\sum_{n>0, p\ge 0} \log(1-q^{n+p+j+\frac{k_b w r_-}{2}} \bar{q}^{p+j+\frac{k_b w r_-}{2}}) 
%\nonumber \\
% & -\log(1-e^{ \pi  k_b w r_+ r_-} q^{p+j} \bar{q}^{p+j}) \, .
-\sum_{n>0, p\ge 0}\log(1- q^{p+j+\frac{k_b w r_-}{2}} \bar{q}^{n+p+j+\frac{k_b w r_-}{2}}) \nonumber \\
&-\sum_{p\ge 0}\log(1- (q\bar q)^{p+j+\frac{k_b w r_-}{2}} )
 \, .
\end{align}
Here we have defined
$q= e^{-2\pi(r_+ + i r_+)}$ and $\bar q = e^{-2\pi (r_+ - i r_-)}$. By  the logic of \cite{Denef:2009kn}, which we follow in  Appendix \ref{QuasiNormalPoles}, identifying the poles of $e^{Z_{\text{saddle}}}$ with the locations of the quasinormal modes, we  propose that there are winding quasinormal modes at frequencies:
\begin{align}
 E_{string} &= \mp L_{string} - i k_b w r_+ r_- -\ii (r_+ \pm ir_-)(2p+2j)
 \label{MatchingModes} \\
 E_{string} &= \mp L_{string} -i k_b w r_+ r_- + \ii (r_+ \pm ir_-)(2p+2j+2 k_bw r_-)
 \label{OtherModes}
  \, .
 \end{align}
It will be interesting to see whether this  treatment of primaries only holds up to closer scrutiny.  Unfortunately, we are unable to provide a similar calculation for descendants at this stage. An independent and more direct argument for the validity of part of the quasinormal winding spectrum is provided in the next subsection.

\subsubsection{A Space-time Description of the Quasinormal Winding Strings}
One way to describe the string modes we identified above  is in terms of a curved doubled field theory, namely a doubled field theory \cite{Hull:2009mi} adapted to our curved, orbifolded group manifold.
Consider generalized scalar wave-functions $\Phi_{double}$ of string states with a dependence on the coordinates $(t,\phi,r)$ of the BTZ metric and on top of that a winding dependence on a dual coordinate $w_D$, $\Phi_{double}=\Phi(t,\phi,r) e^{i w w_D} 
$. We think of the left and right-moving Virasoro constraints as imposing generalized wave operator equations (involving derivatives $\partial_{\phi,t,r}$ but also $\partial_{w_D}$) on the Lorentzian wave-functions $\Phi_{double}$:
\begin{align}
L_0-1 &= \Box + h+N-1 - w(r_+ - r_-) {J}
- \frac{k_b}{4} (r_+-r_-)^2 w^2 \, 
\nonumber \\
\bar{L}_0-1 &= \Box + \bar{h}+\bar{N}-1 - w(r_+ + r_-) \bar{J}
- \frac{k_b}{4} (r_++r_-)^2 w^2 \, ,
\end{align}
where we denoted the string zero modes of the hyperbolic currents  by $J
$ and $\bar{J}
$ -- see \cite{Hemming:2001we,Ashok:2021ffx} -- and the wave-operator $\Box$ equals the quadratic Casimir of the representation in which the ordinary wave-function $\Phi$ transforms.
The wave-function $\Phi$ codes the dependence on the time $t$ and angular coordinate $\phi$ of the string through $(E_{string},L_{string})$ where $L_{string}$ is related to the $-i \partial_\phi$
eigenvalue $L_\phi$ through \cite{Hemming:2001we}
\be
L_{string} = L_\phi + k_b w r_+ r_- \, .
\ee The time translation charge $E_{string}$ and angular charge $L_\phi$ are the ones that appear in the quadratic wave-operator $\Box$ that involves time and angular derivatives. 
The current zero modes are related to the string energy and angular momentum  by the relations \cite{Hemming:2001we}:\footnote{Our notation matches \cite{Ashok:2021ffx}.}
\begin{align}
    (r_+-r_-)J
    &= \frac{E_{string}
    -L_{string} + k_b w r_+ r_-
    }{2}
\nonumber \\
(r_++r_-)\bar{J}
&= \frac{E_{string}
+ L_{string}-k_b w r_+ r_-}{2}
 ~.
\end{align}
A given wave-function $\Phi_{double}$ will satisfy the physical state conditions if on the one hand the wave-function $\Phi$ satisfies the ordinary wave-equation $\Box \Phi=-j(j-1)/k$, and on the other hand, the string energy and angular momentum are chosen such that the on-shell conditions $L_0=1=\bar{L}_0$ are satisfied. 

We choose the boundary conditions on the wave-function $\Phi$ in a discrete representation $j$ to be  in-going at the horizon and without energy loss at infinity. The modes satisfying these boundary conditions are the quasinormal modes. 
The analysis of these wave-functions in the BTZ black hole background is well-known \cite{Birmingham:2001pj}.  There is a spectrum of quasinormal modes that satisfy one of the two  equations \cite{Birmingham:2001pj}: 
\begin{align}
i J 
= j+p~, \qquad  
-i \bar{J} 
=  j+\bar{p} \, \label{TwoConditions}
\end{align}
where $p \ge 0$ or $\bar{p} \ge 0$ is a positive integer. We have used here that it is the string eigenvalues that determine the time and angular dependence of the wave-functions. For instance, if we assume the first equation in (\ref{TwoConditions}) is satisfied, we find the quasinormal frequencies:
\begin{align}
E_{string} &= L_{string} - k_b w r_+ r_- -i (r_+-r_-) (2j+2p)
\, ,
\end{align}
where the string angular momentum $L_{string}$ is quantized. This is a stringy modification of the particle quasinormal mode spectrum (\ref{ParticleQNM}). We note that these  modes, derived from a space-time perspective, match a tower (\ref{MatchingModes}) of quasinormal modes that  we found through the world sheet approach (after analytic continuation to the Lorentzian). To reproduce the remaining towers is a challenge, since one needs  to model the spin and angular momentum  of the excitations more accurately in the space-time wave-function $\Phi_{double}$. This requires further development of the curved doubled field theory that we  introduced.

\subsection{The Long String  Poles}
In this second part, we treat the piece of the partition function that is missed by the saddle point analysis of the previous section. These are the divergent contributions that arise from the poles of the $\theta_1$ function in the denominator.\footnote{These were missed because we assumed $s_1$ to be in the open interval $0<s_1<1$ when performing the expansion of the $\theta_1$ function.} Such a pole was analyzed in \cite{Ashok:2021ffx}. Our analysis will be more complete.
 
We start over with  the one loop string amplitude \eqref{ZandZBTZsaddle}: 
\begin{align}
%\label{ZandZBTZ}
%Z &= \int_F \frac{d^2 \tau}{2 \tau_2} Z_{BTZ} Z_{gh} Z_{int}
%\nonumber \\
 Z&=2r_+ \sqrt{k_b-2}\int_F \frac{d^2\tau}{\tau_2^{3/2}} 
\sum_{m,w} \frac{e^{- \pi \frac{k_b}{\tau_2} r_+^2 |m-w \tau|^2+ \frac{2 \pi}{\tau_2} \text{Im}(\bar{U}_{m,w})^2}}{|\theta_1(\bar{U}_{m,w},\tau)|^2} ~ |\eta(\tau)|^4\, Z_{int}
\, .
\end{align}
We are interested in the behaviour of the partition function near its singularities, which occur at the zeroes of the $\theta$-function. We use the product form of the $\theta$-function 
\be 
\frac{\eta^2(\tau)}{\theta_1(\bar U_{m,w}, \tau)} = \frac{q^{\frac{1}{12}}}{2q^{\frac18}\sin\pi \bar U_{m,w}} \prod_{n=1}^{\infty}\frac{(1-q^{n})}{(1-e^{2\pi i \bar U_{m,w}}q^{n})(1-e^{-2\pi i \bar U_{m,w}}q^n)}~.
\ee 
We have  poles at those values of $\tau$ for which:
\begin{equation}
\label{poleconditions}
\bar{U}_{m,w} +  \tau_{pole} \ell + p = 0 \, ,
\quad \text{or} \quad
\tau_{pole} =\frac{p+ m(r_--ir_+)  }{w(r_- -i r_+) -\ell}
\end{equation}
where $p$ and $\ell$ are integers.
The set of poles again transforms as a doublet under SL$(2,\mathbb{Z})$ modular transformations and by the same unfolding logic as before, we can concentrate on the poles with $\ell=0$ in the strip. 
For the case that $\ell=0$, we see that it is the zero of the sine-function that leads to singularities in the partition function. We can approximate this factor by
\begin{align}
\sin \pi \bar U_{m,w} %&\sim \pi( \bar U_{m,w} + p)\cr
&\sim (-1)^p \pi(r_- - i r_+) w (\tau - \tau_{pole})~,
\end{align}
and the poles in the $\tau$-plane are located at 
\be 
\tau_{pole} = \left(\frac{m}{w} + \frac{p r_-}{w(r_-^2+r_+^2)} \right) + i\frac{ p r_+}{w(r_-^2 + r_+^2)}~.
\ee
We concentrate on the modes with strictly positive winding $w$, such that for $\tau_2>0$, we only have poles with $p\ge 1$.
The contribution from this pole to the $\tau$-integral can be evaluated and we find a logarithmic divergence:
\begin{align}
    \frac{1}{\pi^2 w^2 (r_+^2 + r_-^2)}\int\frac{ d^2\tau}{|\tau-\tau_{pole}|^2} =  \frac{2}{\pi w^2 (r_+^2 + r_-^2)} \log\epsilon~. 
\end{align}
Here, we have defined $\tau - \tau_{pole} = \epsilon$. 
We now evaluate the leading term in the partition function near the pole in the $\tau$-plane, that is the coefficient of the logarithmically divergent term.
For starters we see that the non-trivial exponent simplifies drastically
\be 
- \pi \frac{k_b}{\tau_2} r_+^2 |m-w \tau|^2+ \frac{2 \pi}{\tau_2} \text{Im}(\bar{U}_{m,w})^2 =-\pi k_b p r_+ w~. 
\ee
The contribution from the ghost, internal conformal field theory and rest of the $\theta$-function can again be succinctly expressed  in a $q$-expansion:
\begin{align}
    (q\bar q)^{\frac{1}{4(k-2)} -1}\sum_{h,N} d_{h,N}\, q^{h+N}\bar q^{\bar h + \bar N}~. 
\end{align}
The overall exponent is  fixed by the central charges of the world sheet conformal field theories and $d_{h,N}$ denotes the degeneracies of states with the given quantum numbers. In this expression we substitute $\tau =\tau_{pole}$ to obtain the leading contribution to the partition function. Collecting the various pieces, we find that 
\begin{align}
    Z_{pole} &= \frac{1}{\pi}\,  \log\epsilon\, \sum_{m,w,p}\sum_{h,N} d_{h,N}~\sqrt{ \frac{(k_b-2) (r_+^2+r_-^2) }{p^3 r_+ w}} e^{2\pi i \frac{m}{w}(h+N - \bar h -\bar N)}\cr
    &\hspace{1cm}\times e^{-2\pi p \left( \frac{k_b w r_+}{2} + \frac{1}{w(r_+ + i r_-)}(h + N - 1 + \frac{1}{4(k_b-2)}) + \frac{1}{w(r_+ - i r_-)}(\bar h + \bar N - 1 + \frac{1}{4(k_b-2)}) \right) }  \, .
\end{align}
We now introduce the $s$-integral:
\be 
\int_{-\infty}^{\infty}ds~e^{-\frac{2\pi p}{w}\frac{2r_+}{r_+^2 + r_-^2} \frac{s^2}{k_b-2} } = \frac12 \sqrt{\frac{w(r_+^2 + r_-^2)(k_b-2) }{pr_+} }~. 
\ee
The pole contribution is thereby rewritten as 
\begin{align}
    Z_{pole} &= \frac{2}{\pi}\log\epsilon  \int_{-\infty}^{\infty} ds \sum_{m,w,p}\sum_{h,N}d_{h,N}~ \frac{1}{pw} ~  e^{2\pi i \frac{m}{w}(h+N - \bar h -\bar N)}\cr
    &\hspace{1cm} e^{-2\pi p \left( \frac{k_b w r_+}{2} +  \frac{1}{w(r_+ + i r_-)}(\frac{s^2 + \frac 14}{k_b-2}+ h + N - 1 ) + \frac{1}{w(r_+ - i r_-)}(\frac{s^2 + \frac 14}{k_b-2}+\bar h + \bar N - 1 ) \right) }  \, .
\end{align}
The sum over the integer $m$, where $-1/2 < m/w \le 1/2$ for the pole to lie in the strip,  implies that $h+N-\bar{h}-\bar{N}$  is an integer multiple of $w$. This exponential factor corresponds to the insertion of the operator $e^{2 \pi i m L}$ in the amplitude. The sum over $m$ imposes the angular momentum constraint and adds a factor of $w$, which implies that the pole contribution takes the logarithmic form
\begin{align} 
 Z_{pole} &= \frac{2}{\pi}\log\epsilon  \int_{-\infty}^{\infty} ds \sum_{w,h,N}d_{h,N} \sum_{p=1}^{\infty} 
 \frac{1}{p}e^{-2\pi p \left( \frac{k_b w r_+}{2} + \frac{1}{w(r_+ + i r_-)}(\frac{s^2 + \frac 14}{k_b-2}+ h + N - 1 ) + \frac{1}{w(r_+ - i r_-)}(\frac{s^2 + \frac 14}{k_b-2}+\bar h + \bar N - 1 ) \right) }  \nonumber \\
 &= \frac{2}{\pi}\log\epsilon \int_{-\infty}^{\infty} ds \sum_{w,h,N}d_{h,N}~ \log \big(1- e^{-2\pi f(s)} \big) ~.
 \label{Zpolefinal}
\end{align}
In order to write the second equality, we have recognized the sum over the positive integers $p$ as a sum over multi-long-string contributions that generates a logarithm. We    identify the exponent $f(s)$ associated to a single long string with radial momentum $s$ as
\begin{align} 
f(s) =&\phantom{+} \frac{k_b w (r_++i r_-)}{4} + \frac{1}{w(r_+ + i r_-)}\left(\frac{s^2 + \frac 14}{k_b-2}+ h + N - 1 \right) \cr
&+\frac{k_b w (r_+ -i r_-)}{4} + \frac{1}{w(r_+ - i r_-)}\left(\frac{s^2 + \frac 14}{k_b-2}+\bar h + \bar N - 1 \right)~,
%&= J^2+  {\bar J}^2~,
\end{align}
where we need to impose the level matching condition $N+h - \bar N - \bar h = w L_{string}$, with $L_{string}$ being the angular momentum. We recall that for the long strings the left and right worldsheet conformal dimensions are \cite{Ashok:2021ffx}:
\begin{align}
    L_0 &= \frac{s^2+\frac14}{k_b-2} + h + N  + w(r_+ + i r_-) (iJ) + \frac{k_bw^2}{4}(r_++ i r_-)^2 \\
    \bar L_0 &= \frac{s^2+\frac14}{k_b-2} + \bar h + \bar N  + w(r_+ - i r_-) (-i\bar J) + \frac{k_bw^2}{4}(r_+ - i r_-)^2~,
\end{align}
where $(iJ)$ and $(-i\bar J)$ are identified with the zero modes of the  worldsheet currents in the $w$-twisted sector.  
Solving for the on-shell conditions $L_0 = \bar L_0 =1$, we can solve for these zero modes in terms of the internal dimensions and the twist $w$. We thereby recognize the combination that appears in the exponent in \eqref{Zpolefinal} as 
\be 
-2\pi f(s) =  2\pi(iJ + (- i\bar J))~. 
\ee
The current zero modes are  related to the Euclidean energy and angular momentum of the twisted string \cite{Hemming:2001we,Ashok:2021ffx}: 
\begin{align}
    E_{string} &= (r_+ + i r_-) (iJ) + (r_+ - i r_-) (-i\bar J)\\
    L_{string} &=  -(r_+ + i r_-) (iJ)+ (r_+ - i r_-) (-i \bar J) - i k_b w r_+ r_-~.
\end{align}
We thus have the exponent
\begin{align} 
\label{fintermsofEL}
-2\pi f(s) &= \frac{2\pi r_+}{r_+^2+ r_-^2} E_{string} + \frac{2\pi ir_-}{r_+^2 + r_-^2} (L_{string}- \frac{k_b w}{2}J_{BH}) ~.
\end{align}
We   recall the definition of the boundary modular parameter for the BTZ black hole. It is defined as the S-dual of the boundary modular parameter of the thermal $AdS_3$ background, which is given by  \cite{Maldacena:2000kv}
\be 
\tau_{AdS_3} = \frac{1}{2\pi} (\beta\mu + i \beta)~. 
\ee 
Here $\beta$ is the inverse temperature and $\mu$ is the chemical potential for angular momentum in the thermal $AdS_3$ space-time \cite{Maldacena:2000kv}. In order to write down the partition function of the Euclidean BTZ background, recall that we made the substitutions 
\be 
(\frac{\beta}{2\pi}, \frac{\beta \mu}{2\pi}) = (r_+, r_-)~.  
\ee 
Thus, the boundary modular parameter corresponding to the Euclidean BTZ background is given by 
\be 
\tau_{BTZ} = -\frac{1}{\tau_{AdS_3}} = -\frac{1}{r_- + i r_+}  = -\frac{r_- - i r_+}{r_+^2 + r_-^2}~.
\ee
Putting all this together, one can rewrite \eqref{fintermsofEL} in terms of the boundary modular parameter of the Euclidean BTZ background and interpret the pole contribution to the partition function as the trace over the multiply wound long strings: 
\begin{equation}
Z_{pole}= \frac{2}{\pi} \log \epsilon \, 
\Tr_{\text{multi-long-strings in BTZ}} ~
%e^{-2 \pi H + 2 \pi i L } 
\left[q_{BTZ}^{{\cal L}_0}~ \bar q_{BTZ}^{\overline {\cal L}_0} \right]~,
\label{LongStringPolesContribution}
\end{equation}
where $q_{BTZ} = e^{2\pi i \tau_{BTZ}}$, and we have defined
\be 
\label{spacetimeL0}
{\cal L}_0 = -\frac{E_{string}-{\widetilde L}_{string}}{2}~, \qquad \overline{\cal L}_0 = -\frac{E_{string} + {\widetilde L}_{string}}{2}~,
\ee
with $\widetilde{L}_{string} = L_{string} - \frac{k_b w}{2} J_{BH}$. 

It is an interesting challenge to interpret our final formula  (\ref{LongStringPolesContribution}), for multiple reasons. Firstly, the trace has Hamiltonian insertions ${\cal L}_0$ that have the opposite sign relation to the energy of states. We recall that long strings in the BTZ background have negative energy, as was confirmed by a probe string calculation in  \cite{Ashok:2021ffx}. Our definition of the operator  ${\cal L}_0+\overline{\cal L}_0$ in \eqref{spacetimeL0} leads it to have positive eigenvalues as one may expect of a boundary scaling operator. Secondly, we note that there is no manifest invariance under the T-transformation $\tau_{BTZ} \rightarrow \tau_{BTZ}+1$ in our final result. We explain in Appendix \ref{ModularEquivalenceClasses} that this invariance is lacking when we concentrate on the contribution of a single BTZ black hole background and that it can be be restored by summing over contributions from a class of background geometries. 

A  naive argument would run that the long strings reach infinity and we may therefore  expect to be able to write the pole contribution $Z_{pole}$ to the one-loop integrand as a trace  in the boundary conformal field theory with modular parameter $\tau_{BTZ}$. We note that our final expression takes this form with both a  modification in the sign of the energy operator and in the twist operator $L_{string}$ which is shifted by a contribution proportional to the winding and the black hole angular momentum.\footnote{At zero angular momentum $J_{BH}$ or zero winding, the second point is moot.} However, we note that a winding string reaching asymptotic infinity has a drastic influence on the boundary conformal field theory -- it even changes the central charge of the boundary conformal field theory. Thus, it is at least intriguing that we reach a final expression (\ref{LongStringPolesContribution}) that is  close to an ordinary conformal field theory trace.

\section{Conclusions}
\label{Conclusions}
In this paper, we studied the path integral derivation of the Hilbert space of a particle on the non-compact universal cover $\widetilde{G}$ of SL$(2,\mathbb{R})$. The group $\widetilde{G}$ doubles as the $AdS_3$ space-time geometry and we described how to twist and orbifold the particle partition function on the group, giving a direct algebraic handle on the particle partition sum in thermal $AdS_3$ and BTZ black hole backgrounds. Indeed, we were able to describe these partition sums via an elliptic and hyperbolic orbifold procedure respectively. We thus showed that the BTZ particle partition function affords not only a Euclidean but also a  Lorentzian hyperbolic orbifold interpretation, as well as a description in terms of a sum over quasinormal modes. 

We exploited the lessons we learnt in the context of the first quantized particle to advance our understanding of the string theory spectrum in the BTZ black hole background with NSNS flux. We wrote the one loop vacuum amplitude as a sum over multi-string contributions corresponding to a single string spectrum that we were able to identify. Firstly, we found discrete winding quasinormal modes in a saddle point approximation to the partition function and described their interpretation in space-time. Secondly, we identified divergent pole contributions that correspond to winding long strings in the spectrum of string theory on the black hole background. We thus confirmed the identification of contributions from \cite{Ashok:2021ffx} and put them on a considerably firmer footing. 

While we believe that our results constitute a significant step forward, more work is needed. While hyperbolic characters of groups with sl$(2,\mathbb{R})$ algebra are well-understood, the hyperbolic characters of the affine sl$(2,\mathbb{R})$ algebra deserve further study. These can serve the Lorentzian and algebraic interpretation of the partition function well. Indeed, we have made several approximations in our claim that we identified the two main contributions to the Euclidean partition function, namely an approximation of the integrand near poles and the saddle point approximation to  the discrete contribution. While these contributions are understood to be neatly complementary, they are not as closely intertwined yet as in their (thermal) $AdS_3$ counterpart. A better understanding of the state space origin of affine hyperbolic characters could help in closing this gap. Secondly, the deceptively simple long string pole contribution poses the challenge to provide a  transparent interpretation.   
% Interesting physics in the space of boundary conformal field theories at varying central charge may be coded in our final result -- it is  a  worthy challenge to provide a transparent interpretation  for the long string pole contribution to the one loop amplitude.

More generally, we hope that our understanding of a quantum theory of gravity on black hole backgrounds, both in terms of the perturbative and non-perturbative properties of the spectrum and interactions may advance further, whether through speculative conjectures or laying bricks one by one. 

\section*{Acknowledgements}
 Part of this work was done while S.A. was visiting the Dipartimento di Fisica, Universit\`a di Torino, Italy and he would like to thank the high energy physics group for their warm hospitality.  We would like to thank our colleagues for creating a stimulating work environment.

\appendix

\section{\texorpdfstring{ An Elliptic Orbifold of $AdS_3$}{ An Elliptic Orbifold of AdS}}
\label{AdS3}
We perform an elliptic orbifold on the partition function for a particle on $AdS_3 = \widetilde{G}$. This is a group theoretic procedure to compactify time. Naive analytic continuation relates the closed time $AdS_3$ manifold to Euclidean thermal $AdS_3$. One may therefore expect that the partition functions on these backgrounds agree. In this case, in which the (thermal) circle is topologically non-trivial, the analytic continuation  proceeds  smoothly.
\subsection{\texorpdfstring{Compactifying Time }{ Compactifying Time }}
We derive the partition function for a particle on a closed time $AdS_3$ manifold with an arbitrary radius as an orbifold of $AdS_3=\widetilde{G}$. We start  with the twisted covering group partition function:
\begin{align}
Z(h_l,h_r) &=  \int_{1/2}^{\infty} {dj} \, (\chi_{j}^+(h_l) \chi_{j}^+(h_r)
+\chi_{j}^-(h_l) \chi_{j}^-(h_r))
+ \int_0^\infty ds \int_0^1 d \epsilon \, \chi_{s,\epsilon}^c (h_l)
\chi_{s,\epsilon}^c(h_r) \, ,
\end{align}
with left and right Cartan twists $(h_l,h_r)$. 
To compactify the Lorentzian $AdS_3$ time direction,  we choose to sum over powers of the elliptic projection operator generator $(h_l,h_r)=(e^{i \beta_L^{Lor} t_{\text{ell}}},e^{i\beta_R^{Lor} t_{\text{ell}}})$.  We allow for a different left-right elliptic twist to not only compactify time, but also introduce a fugacity for the $AdS_3$ angular momentum. We find the orbifold partition function:
\begin{align}
Z^{particle}_{El.O. AdS_3} &=   \sum_{n \in \mathbb{Z}} \int_{1/2}^{\infty} {dj} (\chi_{j}^+(e^{in \beta_L^{Lor}t_{\text{ell}}}) \chi_{j}^+(e^{in \beta_R^{Lor}t_{\text{ell}}})
+\chi_{j}^-(e^{in \beta_L^{Lor}t_{\text{ell}}}) \chi_{j}^-(e^{in \beta_R^{Lor}t_{\text{ell}}}))
\, .
\end{align}
We used the expressions for the sl$(2,\mathbb{R})$ characters evaluated on elliptic group elements (see e.g. \cite{Ashok:2022thd}). The continuous characters are zero up to a delta-function contribution which we ignore.   To obtain a more explicit expression for the partition sum, we can write out the discrete characters  for elliptic group elements:
\begin{align}
Z^{particle}_{El.O. AdS_3} &=   \sum_{n \in \mathbb{Z}} \int_{1/2}^{\infty} {dj} 
\left\{
\frac{e^{inj \beta_L^{Lor}}}{1-e^{in \beta_L^{Lor}}} \frac{e^{inj \beta_R^{Lor}}}{1-e^{in \beta_R^{Lor}}} 
+\frac{e^{-inj \beta_L^{Lor}}}{1-e^{-in \beta_L^{Lor}}} \frac{e^{-inj \beta_R^{Lor}}}{1-e^{-in \beta_R^{Lor}}} 
\right\}
\, .
\end{align}
If we view our model as a first quantized particle model on the group manifold and introduce an einbein modulus in the one-loop integral, we find:
\begin{multline}
Z_{1-loop}^{particle}=
\int_0^\infty \frac{dt}{2t}
   \sum_{n \in \mathbb{Z}} \int_{1/2}^{\infty} {dj}  
\left\{ \frac{e^{inj \beta_L^{Lor}}}{1-e^{in \beta_L^{Lor}}} \frac{e^{inj \beta_R^{Lor}}}{1-e^{in \beta_R^{Lor}}} \right. \\
\left. +\frac{e^{-inj \beta_L^{Lor}}}{1-e^{-in \beta_L^{Lor}}} \frac{e^{-inj \beta_R^{Lor}}}{1-e^{-in \beta_R^{Lor}}} 
\right\}e^{-it j(j-1)+it \frac{l^2m^2}{4}} ~.
\end{multline}
To relate the  Lorentzian orbifold result to the Euclidean thermal partition function, we rewrite the expression in terms of the quantity $j-1/2$:
\begin{multline}
Z^{particle}=
\int_0^\infty \frac{dt}{2t} \sum_{n \in \mathbb{Z}} \int_{1/2}^{\infty} {dj} 
\left\{
\frac{e^{in (j-\frac12) \beta_L^{Lor}}}{-2 i \sin (n \beta_L^{Lor})} \frac{e^{in (j-\frac12) \beta_R^{Lor}}}{-2 i \sin (n \beta_R^{Lor})} \right.
\\
\left.+\frac{e^{-in(j-\frac12) \beta_L^{Lor}}}{2i \sin (n \beta_L^{Lor})} \frac{e^{-in(j-\frac12) \beta_R^{Lor}}}{2i \sin (n \beta_R^{Lor})} \right\}
 e^{-it (j-\frac12)^2+i \frac{t}{4}+it \frac{l^2m^2}{4}} \, .
\end{multline}
The $n=0$ term corresponds to a $\beta^{Lor}_{L,R}$ independent divergence associated to a renormalization of the cosmological constant -- we  set $n \neq 0$ from now on. Moreover, as an expression of $j-1/2$ the integrand is even (for each value of $n$), and we can unfold the $j-1/2$ integral into an integral from $-\infty$ to $+\infty$. We find, with $j=1/2+p$:
\begin{align}
Z^{particle}&= -
\int_0^\infty \frac{dt}{8t}
  \sum_{n \neq 0} \int_{-\infty}^{\infty} {dp} 
\frac{e^{in p \beta_L^{Lor}}}{\sin (n \beta_L^{Lor})} \frac{e^{in p \beta_R^{Lor}}}{\sin (n \beta_R^{Lor})}
 e^{-it p^2+i\frac{t}{4}+it \frac{l^2m^2}{4}} \, .
\end{align}
After performing the Gaussian $p$ integral and analytically continuing in the world sheet time $t$, we recognize the Euclidean heat kernel expression -- see e.g. \cite{Giombi:2008vd} or equation (\ref{Projection}) -- up to an overall constant of which we did not keep track.

\subsection{Summary}

We have travelled a  smooth path from the Lorentzian orbifold to the Euclidean partition function. Note that this happened despite the fact that in the Lorentzian, all on-shell states are in discrete representations, while in the Euclidean, all unitary representations are in the continuous representations. In the bulk of the paper, we examine a hyperbolic orbifold in which the thermal (Lorentzian time) circle is topologically trivial.

\section{Hyperbolic Affine Characters}
\label{HyperbolicAffineCharacters}

In this short Appendix, we make a simple point about the proposed hyperbolic affine characters. Firstly, we recall that discrete representations of sl$(2,\mathbb{R})$ with lowest weight state characterized by $j$ have a hyperbolic character -- see e.g. \cite{Ashok:2022thd} --:
\be
\chi_j^+ (e^{t \, t_{\text{hyp}}}) = \frac{ e^{-(2j-1)t}}{|\sinh t|} \, .
\ee
Strictly speaking, this is a distribution on the space of functions on the group. 
We propose that a hyperbolic  character of the corresponding Verma current algebra module is given by:
\be
\hat{\chi}_j^+ (e^{t \, t_{\text{hyp}}}) = \frac{ e^{-(2j-\frac{1}{2})t}}{|\sinh t|} 
\frac{1}{\prod_{n=1}^\infty (1-q^n)(1-e^{2t} q^n)(1-e^{-2t} q^n)}
\, . 
\label{PropAffineCharacter}
\ee
The logic is simple. We only illustrate it at the first level and hope the sequel will be clear. Firstly, consider the lowest weight representation of the current algebra built on the primary representation of spin $j$. Consider it in an elliptic basis. Next, contemplate the first level descendants of the representation, corresponding to the coefficient of the power $q$ in the  expansion of the  character. In the elliptic basis, it is clear how to decompose the tensor product of the discrete lowest weight representation $D_j^+$ of the current algebra primary and the adjoint representation of sl$(2,\mathbb{R})$ made up of the current components $J^{\pm,3}_{-1}$. We obtain representations $D_j^+ \otimes \text{adj} = D_{j-1}^+ \oplus D_j^+ \oplus D_{j+1}^+$. This statement is basis independent. Therefore the character of the representation of the global sl$(2,\mathbb{R})$ representation at this level is necessarily $\chi_{j-1}^+ + \chi_{j}^+ + \chi_{j+1}^+$. For a hyperbolic group element, this character  evaluates to the coefficient of $q$ in the proposed affine character (\ref{PropAffineCharacter}). In the case of the affine continuous character used in the bulk of the paper, the same derivation holds true. The resulting characters are characters of (often non-unitary) continuous  representations. The result of this reasoning is stated in equation (\ref{AffineContinuousCharacter}).

\section{Derivation of the Quasinormal Winding Frequencies}
\label{QuasiNormalPoles}
In this Appendix, we start from the partition sum (\ref{PartitionSum}) and use the method of identifying poles in the Matsubara frequency dependence to surmise quasinormal winding mode frequencies \cite{Denef:2009kn}. The technical details mostly follow \cite{Castro:2017mfj, Keeler:2018lza} but there are some differences due to the shift of the angular momentum of the string by a term proportional to the winding and black hole angular momentum. 

%\subsection{Modes with Positive Winding}

We begin with the saddle point contribution (\ref{WSaddleContribution}) to the partition function from the sector with winding $w$ and rewrite it: 
\begin{align}
    Z^{(w)}_{\text{saddle}}&=\sum_{|m|=1}^{\infty}\sum_{r,\bar r\ge 0}\frac{1}{|m|} e^{-2\pi |m|\left( (r_+ +i r_-)(j+r+\frac{k_b w r_-}{2})+ (r_+ - i r_-)(j+\bar r +\frac{k_b w r_-}{2})  \right)}\cr
    % &= \sum_{|m|=1}^{\infty}  \frac{1}{|m|} (q\bar q)^{|m|(j+\frac{k_b w r_-}{2})}  \sum_{r,\bar{r} \ge 0} q^{|m|r} \bar q^{|m|\bar r}\cr
    &= \sum_{|m|=1}^{\infty}  \frac{1}{|m|} (q\bar q)^{|m|(j+\frac{ k_b w r_-}{2})}  \frac{1}{(1-q^{|m|})(1-\bar q^{|m|})} ~.
    \label{PartitionSum}
\end{align}
In the second line we have defined the space-time nomes
$q= e^{-2\pi(r_+ + i r_+)}$ and $\bar q = e^{-2\pi (r_+ - i r_-)}$.  By manipulating the summation as in  \cite{Keeler:2018lza}, we can rewrite the expression as a sum of three logarithms:
%  \begin{align}
%     Z^{(w)}_{\text{saddle}}&= \sum_{|m|=1}^{\infty}  \frac{1}{|m|} (q\bar q)^{|m|(j+\frac{ k_b w r_-}{2})}\left[ \frac{(1-(q\bar q)^{|m|}) }{(1-q^{|m|})(1-\bar q^{|m|})}\right]\frac{1}{(1-(q\bar q)^{|m|})}\cr
%     &= \sum_{|m|=1}^{\infty}  \frac{1}{|m|} (q\bar q)^{|m|(j+\frac{ k_b w r_-}{2})}\left[ 1+ \sum_{n=1}^{\infty}(q^{|m|n} + \bar q^{|m|n})  \right]\sum_{p=0}^{\infty}(q\bar q)^{|m|p} ~. 
% \end{align}
% Thus, we obtain a sum of three terms, each of which can be written as a logarithm by summing over the variable $|m|$.
\begin{align}
%  \sum_{m=1}^\infty 
%  %e^{ 2\pi m k_b w r_+ r_-}
%  \frac{(q\bar{q})^{m (j+\frac{k_b wr_-}{2})} %\bar{q}^{m(j+\frac{k_b wr_-}{2})} 
%  }{m} 
%  \sum_{r,\bar{r} \ge 0} q^{m r } \bar{q}^{m \bar{r}}
%&= \sum_{m=1}^\infty e^{2 \pi m k_b w r_+ r_-} \frac{q^{m j} \bar{q}^{mj} }{m} \left( \sum_{n=1} (q^{m n }
%+\bar{q}^{m n})+1 \right) \sum_{p \ge 0} (q \bar{q})^{mp} \nonumber \\
% =& -\log(1-e^{ \pi  k_b w r_+ r_-} q^{n+p+j} \bar{q}^{p+j}) -\log(1-e^{2 \pi  k_b w r_+ r_-} q^{p+j} \bar{q}^{n+p+j}) 
% \nonumber \\
Z^{(w)}_{\text{saddle}}=& -\sum_{n>0, p\ge 0} \log(1-q^{n+p+j+\frac{k_b w r_-}{2}} \bar{q}^{p+j+\frac{ k_b w r_-}{2}}) 
-\sum_{n>0, p\ge 0}\log(1-  q^{p+j+\frac{ k_b w r_-}{2}} \bar{q}^{n+p+j+\frac{ k_b w r_-}{2}}) \nonumber \\
&-\sum_{p\ge 0}\log(1-(q\bar q)^{p+j+\frac{ k_b w r_-}{2}} )
 \, .
\end{align}
% Given that 
% \be
% L = -i(r_+ + i r_-)\big(j+r -i \frac{ k_b w}{4}(r_++ i r_-)\big)  - i(r_+ - i r_-) (j+\bar r + i \frac{ k_b w}{4}(r_+ - i r_-))
% \, .
% \ee
% we can identify:
% \begin{align}
% j_l &= -i\big(j -i \frac{ k_b w}{4}(r_++ i r_-)\big)
% \nonumber \\
% j_r &= \pm i (j + i \frac{ k_b w}{4}(r_+ - i r_-)) \, 
% \end{align}
% and (e.g.) $p=r$ (despite some misgivings about the range). 
We exponentiate to rewrite:
\begin{align}
\exp(-2Z^{(w)}_{\text{saddle}})=& \prod_{n>0, p\ge 0} (1-q^{n+p+H} \bar{q}^{p+H})^2 
\prod_{n>0, p\ge 0}(1-  q^{p+H} \bar{q}^{n+p+H})^2 \prod_{p\ge 0}(1-(q\bar q)^{p+H} )^2 \nonumber
 \, .
 \end{align}
 We have defined an effective boundary conformal dimension $H=j + \frac{k_b w r_-}{2}$. We shall denote the exponential by $P_1 P_2P_3$, where each of the $P_i$ corresponds to one of the three infinite products. We now show how this can be equivalently written in a product form in terms of (winding) quasinormal modes.  We define $2\pi T_R = (r_++ i r_-)$ and $2\pi T_L = (r_+ - i r_-)$. We shall consider the first two factors in tandem, and, following  \cite{Denef:2009kn, Castro:2017mfj, Keeler:2018lza}, it is possible to obtain the  expression: 
 % XXX Discuss P3 at the end ? XZZX
 \begin{align}
 P_1P_2&=
 \prod_{n>0, p\ge 0, \ell} \Bigg[\left( \frac{2T_R}{T_R+T_L} \ell+ \frac{4\pi i T_LT_R n}{T_R+T_L} + 2\pi\ii T_R(2p+2H)\right)\cr
 &\hspace{5cm}\times\left( -\frac{2T_L}{T_R+T_L} \ell + \frac{4\pi i T_LT_R n}{T_R+T_L}+ 2\pi\ii T_L(2p+2H)\right)\Bigg] \nonumber \\
 &\prod_{n<0, p\ge 0, \ell} \Bigg[ \left( \frac{2T_R}{T_R+T_L} \ell+ \frac{4\pi i T_LT_R n}{T_R+T_L} - 2\pi\ii T_R(2p+2H)\right) \cr
 &\hspace{5cm}\times \left( -\frac{2T_L}{T_R+T_L} \ell + \frac{4\pi i T_LT_R n}{T_R+T_L}- 2\pi\ii T_L(2p+2H)\right)\Bigg]~.
 \end{align}
 We now recall the Matsubara frequencies \cite{Denef:2009kn, Castro:2017mfj, Keeler:2018lza}
 \be 
 \omega_n = \frac{T_R-T_L}{T_R + T_L} \ell_{\Phi} + \frac{4\pi i T_L T_R}{T_R+T_L}\, n~, 
 \ee 
 where the angular momentum $l_\phi$ associated to translation in the angular coordinate $\phi$ in the BTZ black hole background equals
 $$ 
 \ell_{\Phi} = \ell + i k_b w r_+ r_-~,
 $$ 
 and we have identified $L_{string} = \ell \in \mathbb{Z}$ to be the quantized string angular momentum. The quantum number $n$ is the  quantized Euclidean time momentum. Then,  re-expressing all quantities in terms of the horizon radii $r_{\pm}$, we obtain
%  \begin{align}
% \omega_n +\ell -i k_b w r_+ r_-(\frac{T_R-T_L}{T_R + T_L})&=\frac{2T_R}{T_R+T_L} \ell + \frac{4\pi i T_LT_R n}{T_R+T_L} \\
% \omega_n -\ell + ik_b w r_+r_-(\frac{T_R-T_L}{T_R + T_L})&=-\frac{2T_L}{T_R+T_L} \ell + \frac{4\pi i T_LT_R n}{T_R+T_L} ~. 
% \end{align}
% %
% \begin{align}
%  P_1P_2&=
%  \prod_{n>0, p\ge 0, \ell}\left( \omega_n +\ell -i k_b w r_+ r_-(\frac{T_R-T_L}{T_R + T_L})+ 2\pi\ii T_R(2p+2H)\right)\left( \omega_n -\ell + ik_b w r_+r_-(\frac{T_R-T_L}{T_R + T_L})+ 2\pi\ii T_L(2p+2H)\right) \nonumber \\
%  &\prod_{n<0, p\ge 0, \ell}\left( \omega_n +\ell -i k_b w r_+ r_-(\frac{T_R-T_L}{T_R + T_L})- 2\pi\ii T_R(2p+2H)\right)\left( \omega_n -\ell + ik_b w r_+r_-(\frac{T_R-T_L}{T_R + T_L})- 2\pi\ii T_L(2p+2H)\right)
%  \end{align} 
 
%  \begin{align}
%  T_1T_2&=
%  \prod_{n>0, p\ge 0, \ell}\left( \omega_n +\ell + k_b w  r_-^2 + \ii (r_++ir_-)(2p+2j+k_bw r_-)\right)\left( \omega_n -\ell - k_b w r_-^2+ \ii (r_+-ir_-)(2p+2j+k_bw r_-)\right) \nonumber \\
%  &\prod_{n<0, p\ge 0, \ell}\left( \omega_n +\ell + k_b w  r_-^2- \ii (r_++ir_-)(2p+2j+k_bw r_-)\right)\left( \omega_n -\ell - k_b w r_-^2- \ii (r_+-ir_-)(2p+2j+k_b w r_-)\right)
%  \end{align} 
 \begin{align}
 P_1P_2&=
 \prod_{n>0, p\ge 0, \ell}\Bigg[\left( \omega_n +\ell +
 i k_b w r_+ r_- +\ii (r_++ir_-)(2p+2j)\right) \cr
 &\hspace{3.5cm}\times\left( \omega_n -\ell +i k_b w r_+ r_- + \ii (r_+-ir_-)(2p+2j)\right) \Bigg]\nonumber \\
 &\prod_{n<0, p\ge 0, \ell} \Bigg[\left( \omega_n +\ell +i k_b w r_+ r_- - \ii (r_++ir_-)(2p+2j+2 k_bw r_-)\right)\cr
&\hspace{3.5cm}\times \left( \omega_n -\ell +i k_b w r_+ r_- - \ii (r_+-ir_-)(2p+2j+2 k_b w r_-)\right)\Bigg]
\label{P1P2productfinal}
 \end{align} 
Let us now consider the remaining factor  
\be 
P_3 = \prod_{p\ge 0}(1-(q\bar q)^{p+H} )^2 ~.
\ee 
By following the same sort of logic one can check that we simply obtain the $n=0$ factor in the first product in \eqref{P1P2productfinal}, with $\omega_0  = \frac{T_R-T_L}{T_R+T_L}\ell_{\Phi}$.  
From the product of $P_1$, $P_2$ and $P_3$ one can then read off the quasinormal modes by conjecturally generalizing the principle of \cite{Denef:2009kn} that the quasinormal frequencies arise as zeroes of this expression as a function of the Matsubara frequencies $\omega_n$. For frequencies $\omega_{n\ge0}$, we have energies $E_{string}=\omega$, with 
 \begin{align}
 \omega &= -\ell - i k_b w r_+ r_- -\ii (r_++ir_-)(2p+2j)
 \nonumber \\
 \omega &= +\ell - i k_b w r_+ r_- -\ii (r_+-ir_-)(2p+2j)~,
 \end{align}
while for frequencies with $\omega_{n<0}$, we have the modes
\begin{align}
 \omega &= -\ell -i k_b w r_+ r_- + \ii (r_++ir_-)(2p+2j+2 k_bw r_-)
 \nonumber \\
\omega &= +\ell -i k_b w r_+ r_- +\ii (r_+-ir_-)(2p+2j+2 k_b w r_-)~.
 \end{align}
In Lorentzian conventions, for frequencies  $\omega_{n\ge0}$, we have 
 \begin{align}
  \label{w>0omega>0}
 \omega &= -\ell + k_b w r_+ r_- -\ii (r_+ -r_-)(2p+2j)
 \nonumber \\
 \omega &= +\ell + k_b w r_+ r_- -\ii (r_+ + r_-)(2p+2j)~,
 \end{align}
while for frequencies  $\omega_{n<0}$, we have the modes
\begin{align}
 \label{w>0omega<0}
 \omega 
 &= -\ell +2k_b w r_-^2 - k_b w r_-r_+  + \ii (r_+ - r_-)(2p+2j)\nonumber \\
 %&= -\ell + k_b w r_+r_- - 2k_bwr_-(r_+ - r_-) + \ii (r_+ -r_-)(2p+2j) \nonumber\\
\omega 
%&= +\ell  - k_b w r_-(r_+ + 2r_- ) +\ii (r_+ + r_-)(2p+2j)~. \nonumber \\
&= +\ell -2 k_b w r_-^2  - k_b w r_- r_+  +\ii (r_+ + r_-)(2p+2j)~.
%&= +\ell  + k_b w r_+r_- -2 k_b w r_-( r_++r_-)  +\ii (r_+ + r_-)(2p+2j)~.
 \end{align} 
 These are the winding quasinormal mode frequencies quoted in equations (\ref{MatchingModes}) and (\ref{OtherModes}). We wish to stress that the assumption that the zeroes in the Matsubara frequencies indeed also give the frequencies of fundamental string winding modes could benefit from further corroboration. We provide an independent argument for one of these towers of modes in the bulk of the paper.

\subsubsection*{Symmetries}
We perform a mild check on our proposed spectrum. A $\mathbb{Z}_2$  symmetry of the world sheet action in the BTZ background consists in the combined action of world sheet parity,  $r_- \leftrightarrow -r_-$ as well as the flip $\phi \leftrightarrow -\phi$ of the BTZ angular coordinate $\phi$. The first operation combined with the third leaves the B-field invariant while the second combined with the third leaves the metric invariant. Moreover, the first combined with the second action leaves the orbifold group action invariant.  
On the quantum number $L_{string}=\ell$ this symmetry acts as $L_{string} \leftrightarrow -L_{string}$. 
When we act on our spectrum, we note that $(\ell,r_-)\leftrightarrow (-\ell,-r_-)$ leaves the  set of modes indexed by the integer $w$ invariant.

\section{Modular Equivalence Classes}
\label{ModularEquivalenceClasses}
The solid torus corresponding to Euclidean thermal $AdS_3$ or Euclidean BTZ has a boundary torus with modular parameter $\taust$. As discussed in pedagogical detail in
\cite{Dijkgraaf:2000fq}, the fact that the non-contractible cycle in the solid torus is ambiguous (up to the addition of the contractible cycle) makes for the more accurate statement that a given bulk geometry corresponds to an equivalence class of boundary modular parameters $\taust$. In turn, this implies that the path integral partition function must be invariant under the equivalence relation. 

For the BTZ path integral (\ref{ZBTZ}) under study in the bulk of the paper this implies that it must be invariant under the transformation:
\begin{equation}
r_- \rightarrow r_-+1 \, .
\label{TTransformation}
\end{equation}
Using the elliptic property of the $\theta_1$ function:
\begin{align}
|\theta_1(\bar{U}_{m,w}+m-w \tau,\tau)|^2
&=  |q|^{-w^2} e^{-4 \pi  \text{Im}(\bar{U}_{m,w}) w} \, ,
\end{align}
it is straightforward to varify that the path integral (\ref{ZBTZ}) is indeed invariant under the transformation (\ref{TTransformation}).

We note a crucial consequence of this invariance. For the thermal $AdS_3$ partition function, the invariance (\ref{TTransformation}) translates into the invariance of the partition function under the transformation $\beta \mu \rightarrow \beta \mu + 2 \pi$ where $\beta \mu$ functions as the angular chemical potential. Thus, the partition function is periodic in the angular fugacity due to invariance under the $T$ transformation $\taust \rightarrow \taust + 1$. For the Euclidean BTZ geometry however, the periodicity in the inner horizon radius $r_-$ implies periodicity in the real part of $-1/\taust$. Once more this is a consequence of the thermal circle being topologically non-trivial for thermal $AdS_3$ and topologically trivial for BTZ. To restore periodicity in the real part of $\taust$ for the BTZ partition function, we may sum over bulk geometries that differ by the twist $\taust \rightarrow \taust+1$.

\bibliographystyle{JHEP}

\end{document}